\documentclass[aps,prl,twocolumn,noshowpacs,superscriptaddress]{revtex4-1}
\usepackage{xspace}
\usepackage{bm}
\usepackage{graphicx}
\usepackage{xcolor,url}
\usepackage{amsmath,amssymb,amsfonts,amsthm,physics}
\usepackage[colorlinks=true,linkcolor=blue,anchorcolor=red,citecolor=blue,urlcolor=blue]{hyperref}

\def \L {\mathcal{L}}

\def \H {\mathcal{H}}

\def \K {\hat{\mathcal{K}}}

\def \Z {\mathbb{Z}}

\def \k {\bm{k}}

\begin{document}

\title{The gauge-field extended $k\cdot p$ method and novel topological phases}
\author{L. B. Shao}
%\email[These authors contributed  equally  to this work.]{}
\affiliation{National Laboratory of Solid State Microstructures and Department of Physics, Nanjing University, Nanjing 210093, China}
\affiliation{Collaborative Innovation Center of Advanced Microstructures, Nanjing University, Nanjing 210093, China}
\author{Q. Liu}
%\email[These authors contributed  equally  to this work.]{}
\affiliation{National Laboratory of Solid State Microstructures and Department of Physics, Nanjing University, Nanjing 210093, China}
%\affiliation{Collaborative Innovation Center of Advanced Microstructures, Nanjing University, Nanjing 210093, China}
\author{R. Xiao}
\affiliation{National Laboratory of Solid State Microstructures and Department of Physics, Nanjing University, Nanjing 210093, China}
\author{Shengyuan A. Yang}
%\email[]{shengyuan_yang@sutd.edu.sg}
\affiliation{Research Laboratory for Quantum Materials, Singapore University of Technology and Design, Singapore 487372, Singapore}
\author{Y. X. Zhao}
\email[]{zhaoyx@nju.edu.cn}
\affiliation{National Laboratory of Solid State Microstructures and Department of Physics, Nanjing University, Nanjing 210093, China}
\affiliation{Collaborative Innovation Center of Advanced Microstructures, Nanjing University, Nanjing 210093, China}

%We present a theory for folding the Brillouin zone by doubling the primitive cell of a tight-binding model. Based on the theory, we find the Brillouin-zone folding may overlap Fermi points at separated momenta, and therefore can lead to various topological phases transitions. This is demonstrated by a number of models. The topological phase transition of the famous SSH model is reinterpreted in the framework. Time-reversal invariant spinless tight-binding models coupled with $\Z_2$ gauge fields are studied by the theory, including one on the $2$D rectangular lattice with half flux for each plaquette and another on $3$D graphite lattice with half flux for each vertical rectangular plaquette. A variety of topological phases are realized by topological phase transitions in this class: topological insulators with helical edge modes, $2$D Dirac semimetal with edge flat Fermi arcs, $3$D crystalline topological insulator with surface Dirac points, and second-order nodal-line semimetal with hinge Fermi arcs.

\begin{abstract}
Although topological artificial systems, like acoustic/photonic crystals and cold atoms in optical lattices were initially motivated by simulating topological phases of electronic systems, they have their own unique features such as the spinless time-reversal symmetry and tunable $\Z_2$ gauge fields. Hence, it is fundamentally important to explore new topological phases based on these features. Here, we point out that the $\Z_2$ gauge field leads to two fundamental modifications of the conventional $k\cdot p$ method: (i) The little co-group must include the translations with nontrivial algebraic relations; (ii) The algebraic relations of the little co-group are projectively represented. These give rise to higher-dimensional irreducible representations and therefore highly degenerate Fermi points. Breaking the primitive translations can transform the Fermi points to interesting topological phases.
%We first show that, with certain $\Z_2$ gauge fields, the band structure exhibits the $\pi$-periodicity because of the anti-commutative relations of lattice translators and other spatial symmetries from the AB effect. The $\pi$-periodicity motivates us formally double the unit cells and construct a folded Brillouin zone (BZ). The boundary points of the folded BZ ($\pm \pi/2$ in the original BZ) typically host highly degenerate Fermi points projectively representing the translational symmetry, time-reversal symmetry and other crystalline symmetries. Furthermore, translational-symmetry-breaking dimerizations can transform the Fermi points into unexpected topological phases.
We demonstrate our theory by two models: a rectangular $\pi$-flux model exhibiting graphene-like semimetal phases, and a graphite model with interlayer $\pi$ flux that realizes the real second-order nodal-line semimetal phase with hinge helical modes. Their physical realizations with a general bright-dark mechanism are discussed. Our finding opens a new direction to explore novel topological phases unique to crystalline systems with gauge fields and establishes the approach to analyze these phases.
\end{abstract}

\maketitle

{\color{blue}\textit{Introduction.}}
%During the past decade and a half, topological phases of quantum matter have been one of main topics of physics research. Conceptually, the field started with gapped topological systems, including topological insulators and superconductors, and then evolved into gapless topological systems, including Weyl and Dirac semimetal, nodal-line semimetals and even nodal-surface semimetals, with various symmetries, such as elementary symmetries like time reversal, particle-hole and inversion , and crystalline symmetries, with or without spin-orbit coupling. The influence even goes beyond condensed matter physics, as various topological phases of electrons have been simulated by periodic artificial systems, such as cold atoms, photonic and phononic crystals, electric-circuit arrays, and even mechanical systems.
Symmetry-protected topological phases have been one of the most active fields during the past decade and a half~\cite{Hasan2010,Qi2011,Chiu2016,Bansil2016,Armitage2018}. The influence even goes beyond condensed matter physics. Various topological phases initially proposed for electrons in solids have been simulated by periodic artificial systems, such as cold atoms~\cite{Bloch2014,Zhang2018,Cooper2019}, phononic and photonic crystals~\cite{Lu2014,Wang2015,Zhang2015,Li2018,Ozawa2019,Yang2019,Zeng2020}, electric-circuit arrays~\cite{Imhof2018,Yu2020}, and even mechanical systems~\cite{Prodan2009,Huber2015,Hatsugai2017}. This line of research proves to be especially fruitful, as we have well developed techniques for engineering artificial systems to fine tune the band topology, which is impossible for realistic materials.

Recently, it was realized that artificial systems can also exhibit unique features distinct from their electronic forerunner. First, the excitations of artificial systems can be spinless or of integer spins, while electrons have spin-1/2. Accordingly, they follow the algebra of time-reversal symmetry: $(T)^2=1$, in contrast to $(T)^2=-1$ for electrons with spin-orbit coupling, hence they correspond to different topological classifications~\cite{Schnyder2008,Kitaev2009,zhaoPRL2013,zhaoPRL2016a}. Second, with $T$-invariance, artificial systems have intrinsic $\Z_2$ gauge fields, i.e., the hopping amplitudes are real numbers that can take either positive or negative signs. %Note that the $\Z_2$ gauge field preserves the essential $T$ symmetry of artificial systems.
Actually, engineered $\Z_2$ gauge fields have already been utilized to achieve topological phases such as higher-order topological insulators~\cite{Keil2016,Peterson2018,Serra-Garcia2018,Bloch2019,Mittal2019,Xue2020,Ni2020,Qi2020}. Thus, while lessons from electronic systems are important, to explore unique topological phases for artificial systems, it is crucial to further study mechanisms or theories tied to their own characteristic features.

The standard tool to analyze topological criticality in electronic systems is the renowned $k\cdot p$ method, by which numerous topological semimetals and insulators have been studied~\cite{Zhang2019,Vergniory2019,Tang2019a,Weikang_nodal-surface,Zhi-Ming_Nodalline,Zhi-Ming_no-go,Weikang_higher-order-Dirac,Ziming_higher-order-Dirac}. However, the conventional $k\cdot p$ method is insufficient in the presence of gauge fields, because space groups are now \textit{projectively}, rather than regularly, represented. Particularly, the essential ingredients of the $k\cdot p$ method, namely, the little co-group at a $k$ point in the Brillouin zone (BZ) and the algebraic relations of the group elements, are fundamentally modified.  In this Letter, focusing on the $T$-invariant $\Z_2$ gauge field, we reveal two essential modifications: (i) Lattice translations should be taken into account for little co-groups; (ii) The elements of the little co-group follow $\Z_2$ projective algebraic relations inherited from the space group~\cite{Note_ProjectiveRep}.

Recall that conventionally space groups are regularly represented, and points in the BZ label the irreducible representations (IRREPs) of translations. Hence, for a given $\bm{k}$, the unit translation $L_{\bm{a}_i}$ for a lattice vector $\bm{a}_i$ is represented by a constant $e^{i\bm{k}\cdot\bm{a}_i}$.
%Recall that for the conventional $k\cdot p$ method, lattice translations do not explicitly appear in the little cogroup to formulate the low-energy effective $k\cdot p$ models at high-symmetry points in the Brillouin zone (BZ).
In contrast,  in the presence of $\Z_2$ fluxes, the translations can acquire a nontrivial representation and should be explicitly analyzed in the little co-group.
With the extended $k\cdot p$ method, we show that the two modifications mentioned above can generate highly degenerate Fermi points corresponding to higher dimensional IRREPs of the modified little co-group.
%As we shall show, the projective algebraic relations of the translations with other spatial symmetries may lead to a $\pi$-periodicity in the BZ, and therefore it is more appropriate to work in a folded BZ with formally doubled unit cells. The translation then acts as a nontrivial operator in a little co-group. Furthermore, elements in the little co-group inherits projective algebraic relations from the projectively represented space group, which can generate highly degenerate Fermi points.
Furthermore, breaking of the primitive translation, for instance by certain dimerization, can lead to unexpected novel topological phases.

We demonstrate our theory by two interesting models. First, a graphene-like semimetal phase is realized in an alternatively dimerized rectangular lattice with $\pi$-flux per plaquette, where two bulk Weyl points lead to a flat band of edge zero-modes. Second, we realize a real second-order nodal-loop semimetal on a graphite lattice with interlayer $\pi$ flux and alternative dimerization. The nodal loops have both the first and second Stiefel-Whitney topological charges, which lead to hinge helical modes. The models can be naturally realized by the bright-dark mechanism, a general approach to achieve $\pi$ flux in artificial systems. 
%We show that it also naturally generates the desired dimerization patterns in our models.

%In this Letter, we present a universal mechanism for topological phase transitions in artificial systems with $\Z_2$ gauge fields, based on $T$ symmetry and $\Z_2$ projectively represented crystalline symmetries.

\begin{figure}
	\includegraphics[scale=0.57]{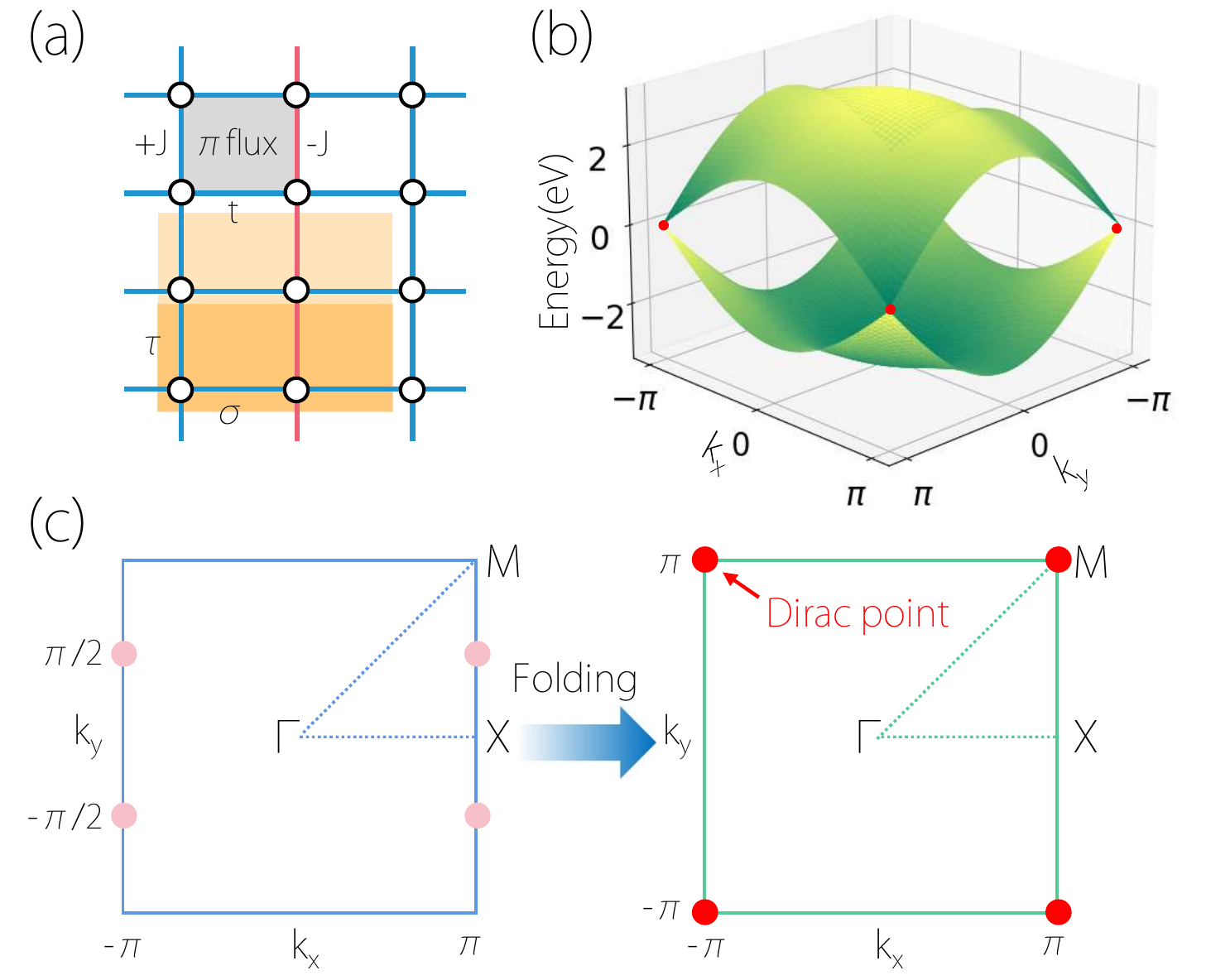}
	\caption{(a) 2D rectangular lattice with $\pi$ flux per plaquette. Red (blue) color marks bonds with a negative (positive) hopping amplitude. (b) A typical folded band structure with a Dirac point at $M$. (c) The Dirac point at $M$ in the folded BZ (right) is formed by two twofold Fermi points at $k_y=\pm\pi/2$ in the unfolded BZ (left). \label{Fig1}}
\end{figure}

{\color{blue}\textit{Extended $k\cdot p$ method}.}  Let us start with a simple case, a lattice translation $L$ and some spatial symmetry $S$, which originally commute with each other.
With $\Z_2$ gauge fields, each closed loop on the lattice will enclose either zero or $\pi$ flux. The key point is that
due to the Aharonov-Bohm effect, the arrangement of $\pi$ flux in the lattice may fundamentally modify the algebraic structure of the crystal symmetries~\cite{Zhao2020-1,Zhao2020-2,Moore2020}.  For example,  under certain $\Z_2$ gauge configuration, $L$ and $S$ could become anti-commutative, i.e.,
\begin{equation}\label{S-L}
[S, L]=0 \quad \Rightarrow\quad \{S,L\}=0,
\end{equation}
when the successive operations $S^{-1}L^{-1}SL$ enclose a $\pi$ flux ($S^{-1}L^{-1}SL=-1$). In this way, the $\Z_2$ projective representation completely changes the symmetry algebra~\cite{Note_Z2-extension}.

An immediate consequence of \eqref{S-L} is that it leads to a $\pi$-periodicity for the energy spectrum. To see this, let $\psi(k)$ be an energy eigenstate with momentum $k\in[-\pi,\pi)$, namely
$L\psi(k)=e^{ik}\psi(k)$. Then, $S\psi(k)$ has the same energy, but with momentum increased by $\pi$, because
\begin{equation}\label{eq2}
  LS\psi(k)=-e^{ik}S\psi(k)=e^{i(k+\pi)}S\psi(k).
\end{equation}
Due to the $\pi$-periodicity, it is appropriate to fold the BZ by formally doubling the unit cell. Another reason is that to construct a BZ compatible with $S$, the eigenvalues of $L^2$ should be used instead of $L$ to define the BZ, since $[L^2,S]=0$. Then, each band on a line in the direction of $L$ and invariant under $S$ has a double degeneracy.

Then, in the folded BZ, although $L^2$ is diagonalized, $L$ generally is not and therefore acquires a nontrivial representation. When we consider a high symmetry point $\k_0$ at $k=0$ or $\pi$ and invariant under $S$, the little co-group $G_{\k_0}$ must contain $L$ as well, since $L$ has projective algebraic relations with other symmetries in the little group. Hence, when represented in terms of generators, we have
\begin{equation}
	G_{\k_0}=\langle L, RT, S,  \cdots \rangle.
\end{equation}
which satisfy the projectively modified algebraic relations:
\begin{equation}
L^2=\pm 1,~ (RT)^2=\pm 1,\\ \{S,L\}=0,~[RT,L]=0,\ \cdots
\end{equation}
Here, $RT$ is a combination of $T$ with a point group element $R$ which leaves $\k_0$ invariant. Whether $L^2=+ 1$ or $-1$ depends on $\k_0$. `$\cdots$' denotes other point group elements and their algebraic relations extended by the $\Z_2$ gauge field. Since $L$ enters the little co-group with nontrivial algebraic relations, higher-dimensional IRREPs typically occur at $\k_0$, leading to highly degenerate points. This argument also shows that momenta $k=\pm \pi/2$ before folding are actually high-symmetry points with degeneracy, since they are mapped to $k=\pi$ of the folded BZ [see Fig.~\ref{Fig1}(c)].
%Located at the folded BZ boundary, this $k=\pi$ point possesses a nontrivial little group $G$: $G$ contains at least $L$ and an anti-unitary symmetry, which is either $T$ or a combination of $T$ with another spatial symmetry, i.e.,

%where $R$ is a unitary element of the space group. Importantly, the translation $L$ enters the little group, which is distinct from the conventional cases.
%This nontrivial group $G$ typically generates highly degenerate (band-crossing) Fermi points [see Fig.], corresponding to multi-dimensional irreducible representations of $G$.

After identifying the little co-group and the projective algebraic relations, we can follow the standard procedure of the $k\cdot p$ analysis. First, we find all IRREPs of the group. Then, for each IRREP, we derive the $k\cdot p$ model $h(\bm{q})$ by implementing the symmetry constraints. Here, the translation $L^{\k_0}$ restricted at $\k_0$ gives the constraint,
\begin{equation}
	[L^{\k_0},h(\bm{q})]=0.
\end{equation}
%The $k\cdot p$ analysis concerns each high-symmetry point of the BZ. 
For tight-binding models on the entire BZ, the BZ folding can be implemented concisely by the approach presented in the Supplemental Material (SM)~\cite{SM}.

{\color{blue}\textit{Graphene-like topological semimetal on rectangular lattice.}}
\begin{figure}
	\includegraphics[scale=0.6]{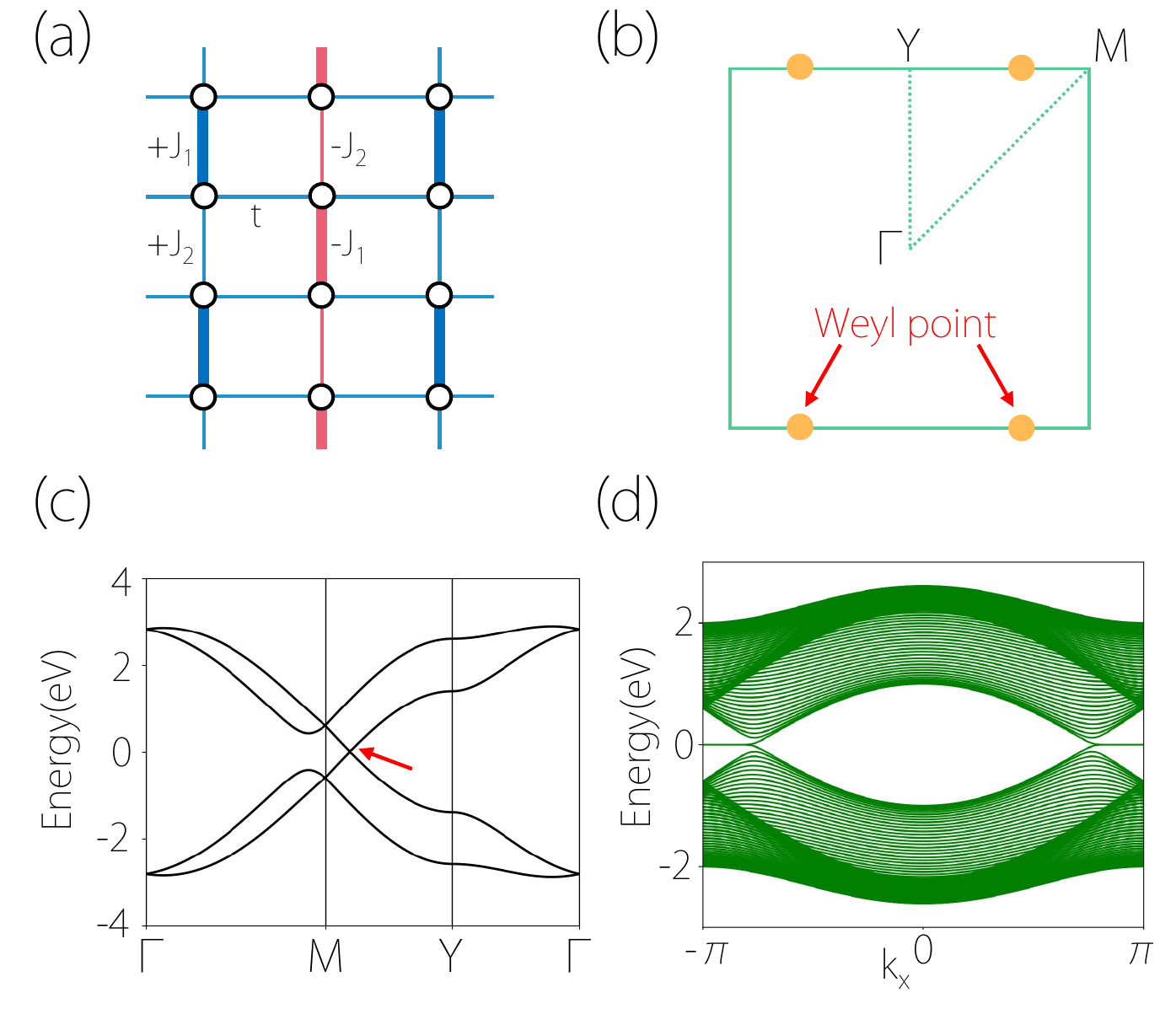}
	\caption{(a) The $\pi$-flux rectangular lattice with a particular dimerization pattern. (b) Under dimerization, the Dirac point at $M$ splits into two Weyl points on the $M$-$Y$ path. (c) shows the corresponding band structure. (d) Spectrum of a slab geometry showing the edge flat band connecting the projection of the two Weyl points. \label{Fig2}}
\end{figure}
Let us consider a 2D lattice as shown in Fig.~\ref{Fig1}(a), where each plaquette has a $\pi$ flux. We will show that independent of model details, the $\Z_2$ gauge field enforces two Fermi points, which after BZ folding overlap into a fourfold degenerate Fermi point at $(\pi,\pi)$ in the folded BZ.

%From the above discussions, we know if there are two Fermi points at $k_0/2$ and $k_0/2+\pi$, respectively, for some $k_0\in[-\pi,\pi)$, then formally doubling the unit cell gives rise to a highly degenerate Fermi point at $k_0$ in the folded BZ as an overlap of the original two Fermi points. We proceed to show that this occurs in a rectangular lattice with $\pi$-flux per plaquette due to the projective representation of the translational symmetries. Because of the projective representation of translational symmetries due to the $\pi$-flux, there are two twofold degenerate Fermi points at $(\pi,\pm \pi/2)$, which overlaps into a fourfold degenerate Dirac point at $(\pi,\pi)$ in the folded BZ.

Let $L_x$ and $L_y$ be the two primitive translation operators for this lattice. Because each plaquette has a $\pi$ flux, a particle moving around a plaquette will acquire a negative sign for its wavefunction, which gives
\begin{equation}\label{L-projective}
\{L_x,L_y\}=0.
\end{equation}
Comparing with \eqref{S-L}, in this case, we have the correspondence that $S=L_x$ and $L=L_y$.
It is natural to consider the set of compatible (mutually commuting) operators $(L_x^2,L_y)$ with eigenvalues $(e^{ik_x},e^{ik_y})$ to specify the BZ.
%They have the common eigenstates $\psi_n(\bm{k})$ with eigenvalues $(\mathcal{E}_n(\bm{k}), e^{ik_x},e^{ik_y})$, respectively.
For an energy eigenstate $\psi_n(\k)$ with energy $\mathcal{E}_n(\bm{k})$, $L_x\psi_n(\k)$ also has energy $\mathcal{E}_n(\bm{k})$. However, $L_x\psi_n(\k)$ locates at $(k_x,k_y+\pi)$ in the BZ, because following Eq.~(\ref{eq2}), $L_yL_x\psi_n(\k)=e^{i(k_y+\pi)}L_x\psi_n(\k)$. Thus, the energy spectrum has the $\pi$-periodicity along $k_y$.

We then double the unit cell  and fold the BZ along $k_y$ [Fig.~\ref{Fig1}(a) and (c)]. The folded BZ is specified by $(L_x^2,L_y^2)$ with eigenvalues $(e^{ik_x},e^{ik_y})$. Note the convention here is that the BZ is always scaled with $2\pi$-periodicity.
%With $L_{x,y}^2\tilde{\psi}_n(\k)=e^{ik_{x,y}}\tilde{\psi}_n(\k)$, it is noted that
Then, the eigenvalues of $L_{x,y}^2$ at $M=(\pi,\pi)$ is $-1$, namely $(L_{x,y}^M)^2=-1$, with $L_{x,y}^M$ the $L_{x,y}$ operators restricted  at $M$.
%Meanwhile, as aforementioned, the $\Z_2$ gauge field preserves the time-reversal symmetry $T=\K$ with $\K$ the complex conjugation.
Hence, at $M$, we need to consider the following little co-group:
\begin{equation}\label{G_M}
	G_M=\langle L_x^M,~L_y^M,~T\rangle,
\end{equation}
with algebraic relations:
\begin{equation}
[T,L_{a}^M]=0,~\{L_x^M,L_y^M\}=0,~(L_{a}^M)^2=-1,~T^2=1,
\end{equation}
where $a=x,y$. It is noteworthy that $T$ is an anti-unitary operator, namely $\{T,i\}=0$ with $i$ the imaginary unit. In the SM, we show that this group has a unique $4$D IRREP~\cite{SM}. Thus, generically there is a fourfold degenerate Dirac point at $M$ in the folded BZ [Fig.~\ref{Fig1}(b)], folded from two twofold Fermi points  at $(\pi,\pm\pi/2)$ in the unfolded BZ [Fig.~\ref{Fig1}(c)].

Above, for simplicity, we choose a small little co-group in \eqref{G_M}. Actually, $M$ is also invariant under mirror symmetries $M_{x,y}$ that inverses $k_{x,y}$, respectively. The unique IRREP can include $M_{x,y}$~\cite{SM}. Moreover, it is noteworthy that without $L_{x,y}^M$, the IRREP will become two dimensional.

The unique IRREP of the little co-group with $M_{x,y}$ gives the $k\cdot p$ model,
\begin{equation}\label{Rectangular_kp}
	h(\bm{k})=\lambda_x k_x \Gamma_2+\lambda_y k_y\Gamma_4 +\mathcal{O}(k^2),
\end{equation}
where the momentum $\bm k$ is measured from $M$, and the Hermitian Dirac matrices are chosen as $\Gamma^1=\tau_0\otimes\sigma_1$, $\Gamma^2=\tau_0\otimes\sigma_2$, $\Gamma^3=\tau_1\otimes\sigma_3$, $\Gamma^4=\tau_2\otimes\sigma_3$, and $\Gamma^5=\tau_3\otimes\sigma_3$, with $\tau$ and $\sigma$ the two sets of the Pauli matrices parameterizing a unit cell.

%For this algebra, it can be shown from the Clifford module theory that there are two irreducible representations, both of which are four-dimensional. Thus, generically there is a fourfold degenerate Dirac point at $M$ in the folded BZ [Fig.~\ref{Fig2}(b)].  Moreover, recalling the $\pi$-periodicity for $k_y$, we see that there are two twofold Fermi points [light red dots in Fig.~\ref{Fig2}(b)] at $(\pi,\pm\pi/2)$, respectively, in the unfolded BZ.

\begin{figure}
	\includegraphics[scale=0.56]{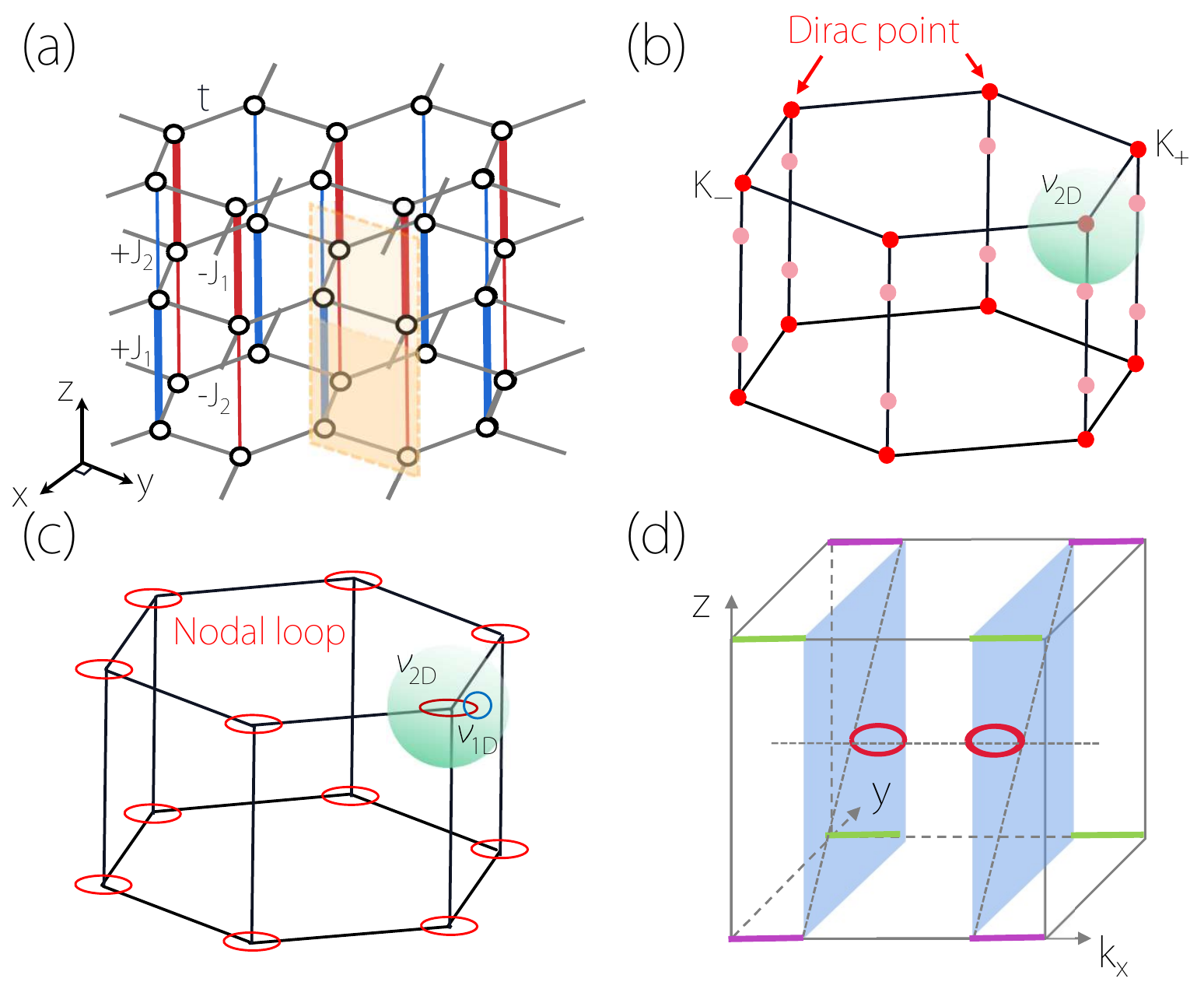}
	\caption{(a) 3D graphite lattice model with flux $\pi$ per rectangular plaquette. Red (blue) color marks bonds with a negative (positive) hopping amplitude. The thickness of the bond indicates a particular dimerization pattern. (b) Without dimerization, there are two Weyl points with opposite chirality at $k_z=\pm\pi/2$ on each vertical edge of the BZ. After doubling the unit cell, they stack into a fourfold real Dirac point with $\nu_\text{2D}$ at a corner of the folded BZ. (c) Under dimerization, each real Dirac point spreads into a real nodal loop with two topological charges. (d) Hinge Fermi arcs appear for a tube-like geometry. Whether it is purple or green inversion pair of edges hosting Fermi arcs depends on the sign of $J_{-}$.  \label{Fig3}}
\end{figure}

We now consider a simple tight-binding model with only the nearest neighbor hopping on this rectangular lattice with $\pi$ flux per plaquette. For the gauge connections in Fig.~\ref{Fig2}(a), a unit cell consists of two sites. Accordingly, the BZ is specified by $L_x^2$ and $L_y$. Then, in the folded BZ specified by $L_x^2$ and $L_y^2$, the Hamiltonian is given by
%the model is given by
%\begin{equation}\label{Half-fluxed-Square-H}
%\H(\k)=2J\cos k_y \sigma_3+t(1+\cos k_x)\sigma_1+t\sin k_x\sigma_2,
%\end{equation}
%where the pseudospin $\sigma$ refers to the two sites as indicated in Fig.
%According to our analysis above, there should be two twofold Fermi points at $(\pi,\pm\pi/2)$, respectively, which is confirmed in Fig.~\ref{Fig2}(b).
%Using Eq.~\eqref{h2}, the Hamiltonian for the formally doubled unit cell is given by
%\begin{equation}\label{Square-double-block}
%\mathcal{H}^{2c}(k_x)=\begin{bmatrix}
%h_0(k_x) & J\sigma_3 + Je^{-\i k_y}\sigma_3\\
%J\sigma_3 + Je^{\i k_y} \sigma_3 & h_0(k_x)
%\end{bmatrix}.
%\end{equation}
%where $h_0(k_x)=t(1+\cos k_x)\sigma_1+t\sin k_x\sigma_2$.
%As we expected, the two twofold Fermi points overlap in the folded BZ to form a fourfold degenerate Dirac point at $M=(\pi,\pi)$.
%It is noteworthy that the explicit expressions for the translational operators in momentum space can only be formulated in the folded BZ as discussed in the SM~\cite{SM}.

%We now introduce the alternative dimerization along the $y$-direction, as illustrated in Fig.\ref{Fig2}(a). The Hamiltonian \eqref{Square-double-block} is now modified to be
\begin{equation}\label{hii}
\begin{split}
\H=\sum_{i=1}^4 f_i(\bm{k})\Gamma^i+g_1(k_y)i\Gamma^3\Gamma^5+g_2(k_y)i\Gamma^4\Gamma^5.
\end{split}
\end{equation}
 The coefficient functions are given by $f_1=t(1+\cos k_x)$, $f_2=t\sin k_x$,  $ f_{3}=J_+(1+\cos k_y) $, $ f_{4}=J_+\sin k_y $, $ g_1=J_-(1-\cos k_y) $, and $ g_2=J_-\sin k_y $, with $ J_{\pm}=(J_1\pm J_2)/2 $. Here, we have introduced an alternating dimerization along $y$ [see Fig.~\ref{Fig2}(a)].
In the absence of the dimerization, we have $J_{-}=0$, and it is straightforward to check that there is a fourfold degenerate Fermi point at $(\pi,\pi)$ following the $k\cdot p$ model \eqref{Rectangular_kp}.
%One may explicitly write Eqs.~\eqref{Square-double-block} and \eqref{hii} into $4\times 4$ matrices to compare them. Only the hopping magnitudes $J$ along the $y$-direction are adjusted to be $J_1$ and $J_2$.

The alternating dimerization with $J_-\ne 0$ breaks $L_x$ and $L_y$, so that the fourfold degenerate Dirac point splits along the $ k_x $ direction into two twofold degenerate Weyl points at $ (\pm K_x,\pi) $ [Fig.~\ref{Fig2}(b, c)]. Each of them has a Berry phase $\pi$ along a circle surrounding it. The quantization of Berry phase is ensured by the $PT$ symmetry with $(PT)^2=1$~\cite{SM}. The two points resemble those in graphene, and they lead to topological edge modes. This is because the two points separate the $k_x$ coordinate $[-\pi, \pi)$ into two segments: $1$D $k_y$-subsystems between them, namely $k_x\in (-\kappa_x,\kappa_x)$, have a trivial Berry phase, whereas those outside, namely  $ k_x\notin[-\kappa_x,\kappa_x] $, have a $\pi$ Berry phase. Hence, there is a flat band of zero-modes connecting the two projected Weyl points at $ \pm \kappa_x $ for an edge along $x$, as shown in Fig.~\ref{Fig2}(d).

%Compared with case (i), the previous $ G $-dressed inversion symmetry represented by $ \hat{\mathsf{P}}=i\tau_2\otimes\sigma_1 $ is now violated when $ J_1\neq J_2 $. Instead, there is another $ G $-dressed off-centered inversion symmetry $ \mathsf{P} $ with the inversion center at middle point of the edge. It can be considered as the combination of $ G $-dressed glide reflection $ G\mathcal{G}_x $ and the mirror symmetry $ M_y $. Therefore, the space-time inversion symmetry $ \mathsf{P}T $ can be represented as
% \begin{equation}\label{PT-1}
% 	\hat{\mathsf{P}}\hat{T}=\left(\begin{array}{cc}
% 		1 & 0 \\ 0 & -e^{ik_y}
% 	\end{array}\right)\otimes\sigma_1\K,
% \end{equation}
% the detail of which can be found in Ref.~\cite{SM}.

{\color{blue}\textit{Real second-order nodal-loop semimetal on graphite lattice.}} Our second example is a graphite lattice with interlayer $\pi$ flux  per rectangular plaquette, as illustrated in Fig.~\ref{Fig3}(a). Here, we take $S$ and $L$ in (\ref{S-L}) to be the mirror symmetry $M_y$ through the $zx$-plane and the primitive translation $L_z$ along $z$, respectively. Because of the flux configuration, they satisfy the anti-commutation relation,
\begin{equation} \label{M-L-projective}
\{M_y,L_z\}=0.
\end{equation}
%Compared with \eqref{L-projective}, $M$ and $L_z$ play the roles of $L_x$ and $L_y$, respectively.
Again, the general analysis below (\ref{S-L}) shows that the band structure has $\pi$-periodicity along $k_z$. Hence, we double the unit cell along $z$ and consider the corners of the folded BZ.  Each corner $K$ of the folded BZ is invariant under the $D_{3}$ group generated by $M$ and $R_{2\pi/3}$ and the combined symmetry $R_{\pi}T$, with $R_{\phi}$ the $\phi$ rotation along the $z$-axis. Hence, we consider the little co-group,
\begin{equation}
	G_K=\langle L_z^K, R_{\pi}T, M_y, R_{2\pi/3} \rangle.
\end{equation}
For the modified algebraic relations of generators, we have
\begin{equation}
	\{L_z^K,M\}=\{L_z^K,R_{\pi}T\}=0,~(L_z^K)^2=-1,
\end{equation}
and the others are ordinary ones~\cite{SM}. In the SM, we show that $G_K$ has two $4$D IRREPs and one $2$D IRREP~\cite{SM}. For the two $4$D IRREPs, the $k\cdot p$ models share the same form of
\begin{equation}\label{Graphite-kp}
\begin{split}
h(\bm{k})=&\lambda_{xy}(k_x\Gamma^1 + k_y\Gamma^2)+\lambda_z k_z\Gamma_4\\
&+\lambda'_{xy}(k_xi\Gamma^1\Gamma^4 + k_yi\Gamma^2\Gamma^4)+\mathcal{O}(k^2).
\end{split}
\end{equation}
In the SM~\cite{SM}, we show that the $\lambda_{xy}'$ term breaks the horizontal mirror $M_z$. With $M_z$ included in the little co-group, $\lambda_{xy}'=0$, and the Dirac point  is folded from two Weyl points with $k_z=\pm \pi/2$ in the unfolded BZ [Fig.~\ref{Fig3}(b)].  Below, we shall see that this Dirac point actually represents a real Dirac point~\cite{Zhao2017}.

%For the graphite lattice quantum states at each corner decompose into a multiplet of the unique $2$D irreducible representation of $D_{3}$. Furthermore, the projective algebraic relation \eqref{M-L-projective} together with $R_{\pi}T$ extends the $2$D irreducible representation into a $4$D irreducible representation of the algebra generated by $M$, $R_{2\pi/3}, L_z$ and $R_{\pi}T$ (see \cite{SM} for technical details). Here, $R_{\pi}$ is the $\pi$-rotation through the $z$-axis. Hence, at each BZ corner there must be a fourfold Fermi point, which results from stacking two twofold Fermi points in the unfolded BZ [Fig.~\ref{Fig3}(b)].

To confirm the general analysis above, we take a tight-binding model with only the nearest neighbor hopping. The Hamiltonian in the folded BZ is given by~\cite{Note_Hamiltonian}
\begin{equation}\label{h-g-ii}
\H(\bm{k})=\sum_{i=1}^{4}\chi_i(\bm{k})\Gamma^i + g_1(k_z) i\Gamma^5\Gamma^4+g_2(k_z) i\Gamma^5\Gamma^3,
\end{equation}
where $\chi_1(\bm{k})+i\chi_2(\bm{k})=t\sum_{i=1}^{3}e^{-i\bm{k}\cdot\bm{a}_i} $ with $ \bm{a}_i $ the three bond vectors for each hexagonal layer, $\chi_{3}(\bm{k})=J_+(1+\cos k_z) $, $\chi_{4}(\bm{k})=J_+\sin k_z $, and $ g_{1,2} $ take the same functional form as in \eqref{hii}. Here, we have added an alternating dimerization pattern, as shown in Fig.~\ref{Fig3}(a).

First, if $J_{-}=0$, the dimerization is switched off, so $L_z$ is preserved and one verifies the fourfold Fermi points at the folded BZ corners. These points are real Dirac points, protected by the $PT$ symmetry. Each is formed from stacking two Weyl points with opposite chirality in the unfolded BZ, and it has the nontrivial $2$D topological charge $\nu_\text{2D}$, which is the real Chern number (also known as the second Stiefel-Whitney number) defined over a sphere surrounding the point [Fig.~\ref{Fig3}(b)]. The topological charge leads to helical Fermi arcs on surfaces parallel to the zigzag direction. The helical arcs can be regarded as resulting from stacking the two chiral Fermi arcs connecting Weyl points in the unfolded BZ~\cite{SM}.

The alternating dimerization with $J_{-}\ne 0$ maintains the $PT$ symmetry. Hence, although the Dirac points are destroyed, the band crossing cannot be completely gapped due to the nontrivial $\nu_\text{2D}$; instead, each Dirac point spreads into a nodal loop normal to the $k_z$-direction [Fig.~\ref{Fig3}(c)]. Note that distinct from ordinary nodal loops, the real nodal loop here has two topological charges $(\nu_\text{1D},\nu_\text{2D})$, where $\nu_\text{1D}$ is the $\pi$-quantized Berry phase for a closed path encircling the loop. It follows that the real nodal-loop semimetal has both drumhead surface states and hinge modes along a pair of inversion-related edges [Fig.~\ref{Fig3}(d)]. The essential physics is revealed in Ref.~\cite{Wang2020}. Here, which pair of inversion-related edges host hinge Fermi arcs is determined by the sign of $J_{-}$, namely the dimerization direction [Fig.~\ref{Fig3}(d)].

{\color{blue}\textit{Discussion and summary}}
%\begin{figure}
%	\includegraphics[scale=0.45]{fig4.pdf}
%	\caption{Model realizations by the dark-bright mechanism. (a, c) The high-energy sites (yellow dots) are inserted to a simple rectangular or graphite lattice to obtain effective negative hopping amplitudes. (b) and (d) are the resulting band structures for (a) and (c), respectively. The low-energy band structures are precisely those for our previously discussed models up to a gauge transformation. \label{Fig4}}
%\end{figure}
Techniques for engineering $\pi$-fluxes or negative hopping amplitudes have been well developed for artificial systems such as photonic/acoustic crystals, electric-circuit arrays, cold atoms, and etc, for which a brief survey has been added in the SM~\cite{SM}.
%In electronic systems, it seems the most straightforward way is to exert magnetic fields. But this is experimentally impractical, since very strong and uniform magnetic fields are required.
Here, we suggest a general approach: When the hopping of a particle between two low-energy sites must go through an intermediate high-energy site, the effective hopping amplitude becomes negative~\cite{Keil2016,SM}. 
%where $t$ is the hopping amplitude between each low-energy site and the high-energy site and $\Delta$ is the site energy difference~\cite{SM}. 
Remarkably, this approach can realize both the gauge flux configuration and the desired dimerization pattern of our models simultaneously~\cite{SM}. 
%As illustrated in Fig.~\ref{Fig4}, our two models can be readily constructed by the ordered insertion of high energy sites~\cite{SM}.

Besides artificial systems, $\Z_2$ gauge fields may also be realized in condensed matter systems. In non-interacting electronic systems, the aforementioned method suggests the ubiquitous existence of $\Z_2$ gauge fields without exerting magnetic fields. Moreover, in quantum spin liquids and Kitaev-type exactly solvable models $\Z_2$ gauge fields emerge in the low-energy effective theories~\cite{Wen-PSG,Kitaev2006}.

It is interesting to consider the generalization of our theory to the case of $U(1)$ gauge fields. For rational flux configurations with denominator $N$, the gauge fields are valued in $\Z_N\subset U(1)$, and we just need to replace $\Z_2$ by $\Z_N$ in our formalism. We note that $\Z_N$ gauge fields with $N>2$ break $T$-invariance. But for irrational fluxes, there is no finite unit cell for any connection configuration. Thus, our formalism is spoiled by the absence of the Brillouin zone. Another aspect is that if the $U(1)$ gauge field for a physical system is tuned to be valued in $\Z_2$, generically there are gauge fluctuations. If the gauge fluctuations are weak, based on our extended $k\cdot p$ method, a low-energy effective theory can be formulated by coupling the $k\cdot p$ model to fluctuations of gauge fields.

In conclusion, we expect our generalized $k\cdot p$ method can be applied to discover numerous unprecedented topological phases in crystalline systems with $T$-invariant $\Z_2$ gauge fields and beyond. With engineerable gauge fields, various artificial systems can be designed and made for realizing the corresponding exotic topological properties.
%: (i) Translations should be taken into account; (ii) The little co-group follows projectively algebraic relations. 
%Beside our models, it can be systematically applied to find more novel topological phases without counterparts in electronic systems.

%{\color{blue}\textit{Acknowledgements.}}The authors acknowledge the support from the National
%Natural Science Foundation of China under Grant (No. 11874201 and No. 11704180).

\begin{acknowledgements}
%This work is supported by National Natural Science Foundation of China (Grants No. 11874201) and the Singapore Ministry of Education AcRF Tier 2 (MOE2019-T2-1-001).
This work is supported by National Natural Science Foundation of China (Grants No. 11874201) and the Singapore Ministry of Education AcRF Tier 2 (MOE2019-T2-1-001). L. B. S. and Q. L. contributed equally to this work.
\end{acknowledgements}

\bibliographystyle{apsrev}
\bibliography{Topo_Folding_Ref}

\begin{thebibliography}{53}
\bibitem[{Not({\natexlab{a}})}]{Note_ProjectiveRep}
\bibinfo{note}{For the conventional cases including both magnetic and
  non-magnetic space groups, the space groups are not projectively represented,
  and projective algebraic relations for a little cogroup occur only for
  nonsymmorphic groups. For the projectively represented space groups, the
  projective algbera for little cogroups can be directly inherited from that of
  the space group. We also note that magnetic groups have essentially different
  algebraic structures from projecive representations of space groups.}
\bibitem[{Not({\natexlab{b}})}]{Note_Z2-extension}
\bibinfo{note}{In mathematical terms, the space group is projectively extended
  by the $\mathbb{Z}_2$ group}.
\bibitem[{Not({\natexlab{c}})}]{Note_Hamiltonian}
\bibinfo{note}{The two Hamiltonian(15) can be converted to the same form of
  (10) by the unitary transformation $U=\Gamma_4\Gamma_5
  \exp(\Gamma_3\Gamma_4\pi/4)$}.
\end{thebibliography}

\newpage
%\begin{appendices}

%\appendix
\onecolumngrid
\renewcommand{\theequation}{S\arabic{equation}}
\setcounter{equation}{0}
\renewcommand{\thefigure}{S\arabic{figure}}
\setcounter{figure}{0}

%\begin{refsection}

\section{Supplemental Material for \\ ``The gauge-field extended $k\cdot p$ method and novel topological phases"}

\section{Folding of the Brillouin zone}

For a 1D chain with $ 2N $ sites, the tight-binding Hamiltonian can be written as
\begin{equation}\label{h1}
	H^c=\sum_{j=0}^{2r}L_{2N}^{j}\otimes A_j+H.c.,
\end{equation}
where $ L_{2N} $, as the generator of the translation group, represents the left-translational operation by one lattice spacing $ c $. $ A_j $ is the operation over the internal degrees of freedom, and $ 2r $ is the hopping range. We assume the dimension of the internal degree is $ n $ and $ A_j $ is an $ n\times n $ matrix. In the momentum space, we can replace $ L_{2N}^{j} $ by $ e^{ikj} $ and the Hamiltonian in Eq.~\eqref{h1} can be rewritten as
\begin{equation}\label{Hck}
	\H^c(k)=\sum_{j=0}^{2r}e^{ikj}A_j+e^{-ikj}A_j^\dagger.
\end{equation}

\begin{figure}[t]
	\centering
	\includegraphics[scale=0.5]{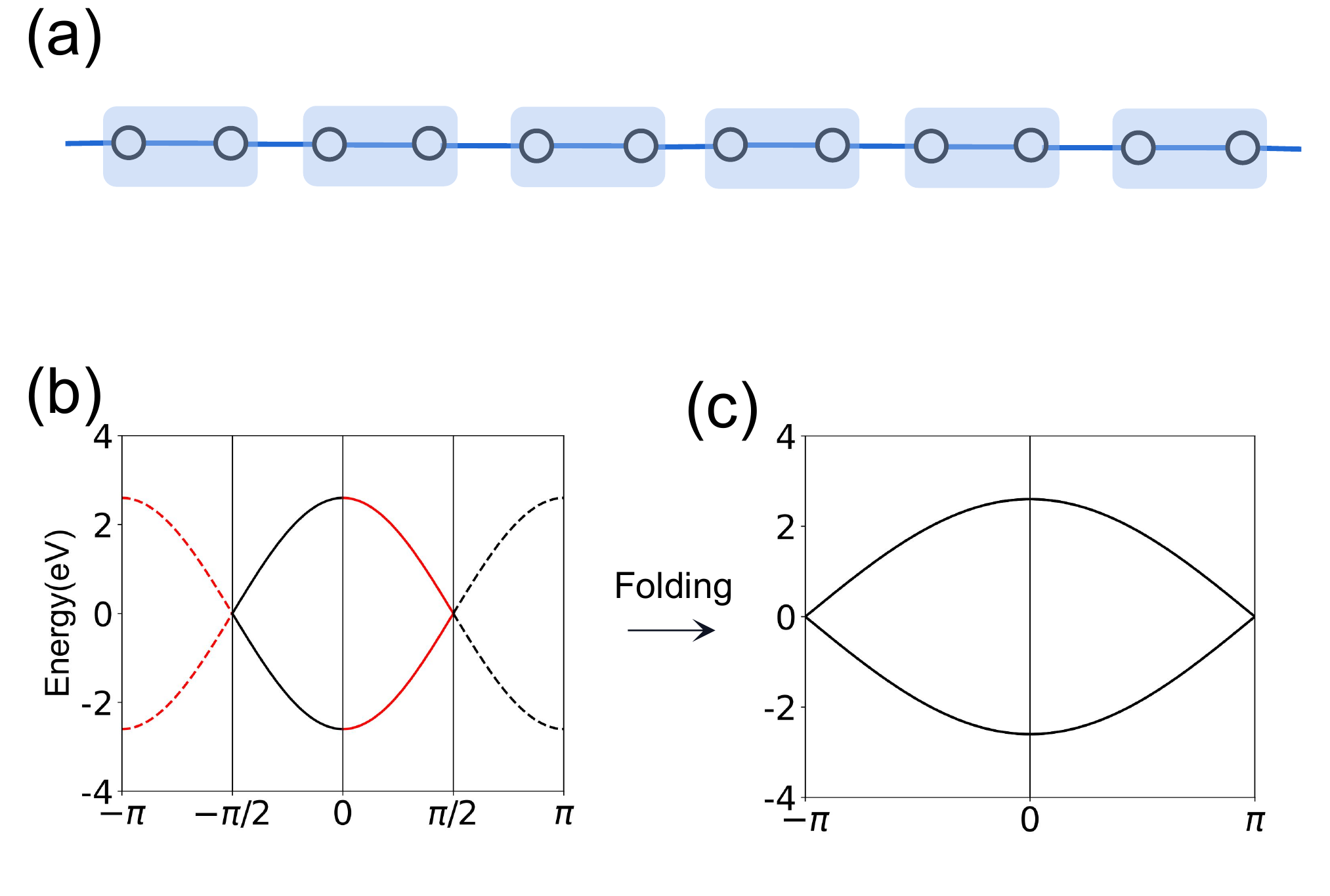}
	\caption{(a) There are two sublattices after folding the BZ. (b) and (c) The folding of band energy bands.\label{figs0}}
\end{figure}

By formally doubling the unit cell, $ i.e. $, folding the Brillouin zone (BZ), there are $ N $ doubled unit cells (DUCs) with two sublattices $ A $ and $ B $ as shown in Fig.~\ref{figs0}. The length of the DUC is doubled as $ 2c $. Under the primitive translation $ L_{2N} $, sublattice $ A $ is translated to sublattice $ B $ of the left DUC, while sublattice $ B $ is translated to the sublattice $ A $ of the same DUC [See Fig.~\ref{figs0}]. Therefore, in the representation of the sublattices, $ L_{2N} $ can be written as
\begin{equation}
	L_{2N}=\begin{pmatrix}
		0 & 1_n\\ L_N & 0
	\end{pmatrix}
\end{equation}
with $ L_N $ as the generator of the translation group after the doubling. Here $ 1_n $ is the $ n\times n $ identity matrix. It further produces $ L_{2N}^2=\tau_0\otimes L_N $, and
\begin{equation}\label{S-decompose}
	L_{2N}^{2j}=\tau_0\otimes L_N^j,\quad L_{2N}^{2j+1}=\tau_+\otimes L_{N}^j+\tau_-\otimes L_{N}^{j+1},
\end{equation}
where $ \tau $'s are the Pauli matrices. Substituting Eq.~\eqref{S-decompose} into Eq.~\eqref{h1}, we obtain
\begin{equation}\label{H2c}
	H^{2c}=\begin{pmatrix}
		B_+ & B_- \\ L_NB_- & B_+
	\end{pmatrix},
\end{equation}
where
\begin{equation}
	\begin{split}
		B_+&=\sum_{j=0}^{r}L_N^j\otimes A_{2j}+h.c.,\\
		B_-&=\sum_{j=0}^{r-1}L_N^j\otimes A_{2j+1}+\sum_{j=1}^{r}L_N^{-j}\otimes A_{2j-1}^\dagger.
	\end{split}
\end{equation}
The Hermiticity of the Hamiltonian is guaranteed by $ B_-^\dagger=L_NB_- $. Replacing $ L_N $ by $ e^{ik} $, the Hamiltonian in Eq.~\eqref{H2c} can be transformed in the momentum space as 
\begin{equation}\label{H2ck}
	\H^{2c}(k)=\begin{pmatrix}
		B_+(k) & B_-(k) \\ e^{ik}B_-(k) & B_+(k)
	\end{pmatrix},
\end{equation}
where
\begin{equation}\label{BB}
	\begin{split}
		B_+(k)&=\sum_{j=0}^{r}e^{ikj} A_{2j}+h.c.,\\
		B_-(k)&=\sum_{j=0}^{r-1}e^{ikj} A_{2j+1}+\sum_{j=1}^{r}e^{-ikj} A_{2j-1}^\dagger.
	\end{split}
\end{equation}
Since formally doubling the unit cell leads to no physical consequence, the primitive translation $ L_{2N} $ is preserved, namely, $ [L_{2N}, H^{2c}] =0$. It is indeed the case for Eq.~\eqref{H2c} since $ H^{2c}=\tau_0\otimes B_++L_{2N}\otimes B_- $. So, $ L_{2N} $ and $ H^{2c} $ can be simultaneously diagonalized. In the momentum space, we can directly replace $ L_{N} $ as $ e^{ik} $, and the primitive translation is obtained as
\begin{equation}\label{L-k}
	\L_{2N}(k)=\begin{pmatrix}
		0 & 1 \\ e^{ik} & 0
	\end{pmatrix}\otimes 1_n.
\end{equation}
It can be diagonalized as
\begin{equation}
	\mathcal{V}(k)\L_{2N}(k)\mathcal{V}^\dagger(k)=\begin{pmatrix}
		e^{\frac{ik}{2}} & 0 \\ 0 & -e^{\frac{ik}{2}}
	\end{pmatrix}\otimes 1_n
\end{equation}
with
\begin{equation}\label{Vk}
	\mathcal{V}(k)=\frac{1}{\sqrt{2}}\begin{pmatrix}
		1 & e^{-\frac{ik}{2}} \\ e^{\frac{ik}{2}}& -1
	\end{pmatrix}\otimes 1_n.
\end{equation}
Using $ \mathcal{V}(k) $ in Eq.~\eqref{Vk} to diagonalize the Hamiltonian in Eq.~\eqref{H2ck}, we have
\begin{equation}\label{HBB}
	\mathcal{V}(k) \mathcal{H}^{2c}(k)\mathcal{V}^\dagger(k)=\begin{pmatrix}
		B_+(k)+e^{\frac{ik}{2}}B_-(k) & 0 \\ 0 & B_+(k)-e^{\frac{ik}{2}}B_-(k)
	\end{pmatrix}.
\end{equation}
Note that the Hamiltonian before doubling the unit cell in Eq.~\eqref{Hck} can be rewritten as
\begin{equation}\label{HBB1}
	\mathcal{H}^c(k )=B_+(2k )+e^{ik }B_-(2k ),
\end{equation}
where Eq.~\eqref{BB} has been used. By comparing Eqs.~\eqref{HBB} and \eqref{HBB1}, we derive the folding relation as
\begin{equation}\label{vhv}
	\mathcal{V}(k) \mathcal{H}^{2c}(k)\mathcal{V}^\dagger(k)=\begin{pmatrix}
		\mathcal{H}^c(\frac{k}{2} ) & 0 \\ 0 & \mathcal{H}^c(\frac{k}{2} +\pi) 
	\end{pmatrix},
\end{equation}
which implies that the points at $ k/2 $ and $ k/2+\pi $ are mapped to the same point at $ k $ after folding the BZ. Note that $ \H^{2c}(k) $ is periodic over the BZ while the right-hand side of Eq.~\eqref{vhv} is not, since $ \mathcal{V}(k) $ is not periodic. 

In the retrospect, we revisit SSH model by the theory of Eq.~\eqref{vhv}. After the folding without the dimerization, the primitive translation symmetry $ L_{2N} $ is preserved such that $ [L_{2N},H^{2c}]=0 $. Since $ L_{2N}^2= L_N^1\otimes 1_2 $, the group generated by $ L_{2N}^2 $ is abelian, which can be represented as $ \L_{2N}^2(k)=e^{ik} $ by Bloch theorem.The same argument can be found in the main text below Eq.~(2). Then, the eigenvalues of $ \L_{2N}(k) $ is obtained as $ \lambda_{\pm}(k)= \pm e^{ik/2} $. In the momentum space, we have $ [\L_{2N}(k),\H^{2c}(k)]=0 $, which leads to two eigenstates $ \ket{k,\lambda_{\pm}(k)} $ corresponding to the eigenenergies $ E_{\pm}(k) $, respectively. The two eigenstates at $ k=0 $ are obtained as $ \ket{0,\pm1} $. After continuously varying $ k=0\rightarrow 2\pi $, $ \ket{0,\pm1} $ evolves into $ \ket{2\pi,\mp} $, respectively. Namely, $ E_{\pm}(0)\rightarrow E_{\mp}(2\pi)$, which means that the primitive translation forces the energy bands crossing at some points in $ [0,2\pi] $, due to the periodicity of the band structure. Consequently, there must be odd number of the band crossing point (Dirac point) in the BZ. Because of the time reversal symmetry, the Dirac points must locate at the time-reversal-invariant point (TRIP). Otherwise, the time reversal symmetry will lead to even number of Dirac points which contradicts the above discussion. We claim the Dirac points at $ k=\pi $, since the bands at the other TRIP $ k=0 $ is always gapped. According to Eq.~\eqref{vhv}, this Dirac point results from two Fermi points at $ k_{\pm}=\pm\pi/2 $ before folding the BZ. By the dimerization, the mass can be introduced to open the gap, which signifies the band inversion.

\section{IRREPs of little cogroups}
In this section, we derive IRREPs of little cogroups in the main text. The general idea is that for each cogroup we first derive all IRREPs of the unitary subgroup, and then group them into complex conjugacy pairs to construct IRREPs of the whole cogroup.

\subsection{The Rectangular Model}
The algebraic structure of the little cogroup is quoted from the main text as
\begin{gather}\label{Rectangular-relations}
	%\begin{split}
	\{L_x^M,L_y^M\}=0,\quad (L^M_x)^2=-1,\quad (L^M_y)^2=-1,\nonumber \\ 
	[L_x^M,T]=[L_y^M,T]=0, \quad T^2=1.
	%\end{split}
\end{gather}
The first line contains the algebraic relations of the unitary subgroup $\Z_2\times\Z_2$, for which the generators are denoted as $l_x$ and $l_y$. The corresponding factor system is given by
\begin{equation}
	\Omega(l_x,l_x)=\Omega(l_y,l_y)=-1,\quad \Omega(l_x,l_y)=-1,\quad \Omega(l_y,l_x)=1,
\end{equation}
noting that $\Omega(l_a,e)=\Omega(e,l_a)=1$ with $e$ the identity of $\Z_2$ and $a=x,y$. It is easy to check that $\Z_2\times\Z_2$ together with the factor system exactly corresponds to the projective algebraic relations in the first line of \eqref{Rectangular-relations}. The group $\Z_2\times\Z_2$ has a unique $2$D projective IRREPs with the factor system. This is because of its order $4$ satisfying $2^2=4$, and the fact that there is no $1$D IRREPs to satisfy the anti-commutation relation $\{L_x^M,L_y^M\}=0$. It is easy to construct matrices for the IRREP:
\begin{equation}
	\pi(l_x)=\begin{bmatrix}
		0 & i\\
		i & 0
	\end{bmatrix},\quad 
	\pi(l_y)=\begin{bmatrix}
		0 & -1\\
		1 & 0
	\end{bmatrix}.
\end{equation}
To include $T$, an anti-unitary element, we simply follow the map from $\mathbb{C}$ to $M_2(\mathbb{R})$,
\begin{equation}
	1\mapsto \begin{bmatrix}
		1 & 0\\
		0 & 1
	\end{bmatrix},\quad 
	i\mapsto \begin{bmatrix}
		0 & -1\\
		1 & 0
	\end{bmatrix}.
\end{equation}
Then, for a complex matrix $A+iB$, 
\begin{equation}\label{Mps2RM}
	A+iB\mapsto \begin{bmatrix}
		1 & 0\\
		0 & 1
	\end{bmatrix}\otimes A+\begin{bmatrix}
		0 & -1\\
		1 & 0 
	\end{bmatrix}\otimes B.
\end{equation}
Thus, the generators in \eqref{Rectangular-relations} are represented by
\begin{equation}\label{L_IRREP}
	L_x^M=\begin{bmatrix}
		0 & 0 & 0 & -1\\
		0 & 0 & -1 & 0\\
		0 & 1 & 0 & 0\\
		1 & 0 & 0 & 0
	\end{bmatrix}\quad 
	L_y^M=\begin{bmatrix}
		0 & -1 & 0 & 0\\
		1 & 0 & 0 & 0\\
		0 & 0 & 0 & -1\\
		0 & 0 & 1 & 0
	\end{bmatrix},\quad T=\begin{bmatrix}
		1 & 0 & 0 & 0\\
		0 & 1 & 0 & 0\\
		0 & 0 & 1 & 0\\
		0 & 0 & 0 & 1
	\end{bmatrix}\K \hat{I}.
\end{equation}
Here $\K$ denotes the complex conjugation. We have added the inversion of momenta $\hat{I}$ into $T$ for the formulation of $k\cdot p$ models later.

Above for the technical simplicity, we only considered translations and time-reversal in the little cogroup of $M$. Actually, we can also include mirror symmetries $M_a$ with $a=x,y$. Here, $M_a$ inverses the $a$ coordinate. The additional algebraic relations are given by
\begin{equation}
	\{M_a,L_b^M\}=0,\quad \{M_x,M_y\}=0,\quad M_x^2=M_y^2=1,\quad 
	[M_a,T]=0
\end{equation}
where $a,b=x,y$. The IRREP \eqref{L_IRREP} can also host the operators $M_{x,y}$, which are given by
\begin{equation}\label{M_IRREP}
	M_x=\begin{bmatrix}
		1 & 0 & 0 & 0\\
		0 & -1 & 0 & 0\\
		0 & 0 & 1 & 0\\
		0 & 0 & 0 & -1
	\end{bmatrix}\hat{I}_x,\quad M_y=\begin{bmatrix}
		0 & 1 & 0 & 0\\
		1 & 0 & 0 & 0\\
		0 & 0 & 0 & -1\\
		0 & 0 & -1 & 0
	\end{bmatrix}\hat{I}_y,
\end{equation}
with $\hat{I}_a$ the inversion of the $k_a$ coordinate. 

Actually, this is the unique IRREP of the projective group algebra. The algebra is equivalent to a real Clifford algebra. The generators are given by
\begin{equation}
	iL_{x},\quad  iL_{y},\quad T,\quad  iT,\quad iM_{x},\quad iM_{y},
\end{equation}
which anti-commute with each other. And the squares of the first four generators are equal to $+1$, and those of the last two are equal to $-1$. Thus, the Clifford algebra is $C^{2,4}$. From the theory of real Clifford algebras, $C^{2,4}$  has a unique IRREP as given above.

To compare with our tight-binding model, it is convenient to use the following matrix representations:
\begin{gather}\label{L_IRREP_2}
	L_x^M=\begin{bmatrix}
		0 & 1 & 0 & 0\\
		-1 & 0 & 0 & 0\\
		0 & 0 & 0 & -1\\
		0 & 0 & 1 & 0
	\end{bmatrix}\quad 
	L_y^M=\begin{bmatrix}
		0 & 0 & 1 & 0\\
		0 & 0 & 0 & 1\\
		-1 & 0 & 0 & 0\\
		0 & -1 & 0 & 0
	\end{bmatrix},\quad T=\begin{bmatrix}
		1 & 0 & 0 & 0\\
		0 & 1 & 0 & 0\\
		0 & 0 & 1 & 0\\
		0 & 0 & 0 & 1
	\end{bmatrix}\K \hat{I},\nonumber \\
	M_x=\begin{bmatrix}
		0 & 1 & 0 & 0\\
		1 & 0 & 0 & 0\\
		0 & 0 & 0 & -1\\
		0 & 0 & -1 & 0
	\end{bmatrix}\hat{I}_x,\quad M_y=\begin{bmatrix}
		0 & 0 & 1 & 0\\
		0 & 0 & 0 & 1\\
		1 & 0 & 0 & 0\\
		0 & 1 & 0 & 0
	\end{bmatrix}\hat{I}_y,
\end{gather}
which are equivalent to \eqref{L_IRREP} and \eqref{M_IRREP} up to a unitary transformation.

\subsection{The Graphite Model}
From the main text we know
\begin{equation}
	\{L_z^K,R_\pi T\}=0.
\end{equation}
It is technically convenient to introduce $\tilde{L}_z^K=iL_z^K$, since 
\begin{equation}
	[\tilde{L}_z^K,R_\pi T]=0.
\end{equation}
Then, the algebraic relations for the unitary subgroup is quoted as
\begin{gather}\label{Graphite_relation}
	%\begin{split}
	\{\tilde{L}_z^K, M\}=[\tilde{L}_z^K,  R_{\frac{2\pi}{3}}]=0,\quad (\tilde{L}_z^K)^2=1 \nonumber\\ (MR_{\frac{2\pi}{3}})^2=1,\quad  (M)^2=(R_{\frac{2\pi}{3}})^3=1.
	%\end{split}
\end{gather}
The projective algebraic relations correspond to the group $D_3\times \mathbb{Z}_2$ with generators $r,m$ and $l$. Here, $r$ and $m$ generate $D_3$, and $l$ generates $\Z_2$. The factor system is specified as
\begin{equation}
	\Omega(l_z,m)=-1,\quad \Omega(m,l_z)=1,
\end{equation}
which correspond to the anti-commutation relation $\{\tilde{L}_z^K, M\}=1$. The other factors are either trivial or can be derived from the factors above. The unitary group $D_3\times \mathbb{Z}_2$ with the factor system has three $2$D IRREPs. This is because of
\begin{equation}
	2^2+2^2+2^2=12
\end{equation}
with $12$ the order of $D_3\times \mathbb{Z}_2$, and the fact that there is no $1$D IRREPs compatible with the anti-commutation relation.

It is easy to construct the three IRREPs. The first one is given by 
\begin{equation}
	\pi_0(l)=\begin{bmatrix}
		0 & 1\\
		1 & 0
	\end{bmatrix},\quad \pi_0(m)=\begin{bmatrix}
		1 & 0\\
		0 & -1
	\end{bmatrix},\quad
	\pi_0(r)=\begin{pmatrix}
		1 & 0\\
		0 & 1
	\end{pmatrix}
\end{equation}
All matrices are already real. Thus, $\pi_0$ gives a $2$D IRREP for the whole little cogroup:
\begin{equation}
	L_z^K=\begin{bmatrix}
		0 & -i\\
		-i & 0
	\end{bmatrix},\quad 
	M=\begin{bmatrix}
		1 & 0\\
		0 & -1
	\end{bmatrix}\hat{I}_{y},\quad 
	R=\begin{bmatrix}
		1 & 0\\
		0 & 1
	\end{bmatrix}\hat{R}^z_{\frac{4\pi}{3}},\quad 
	R_\pi T=\begin{bmatrix}
		1 & 0\\
		0 & 1
	\end{bmatrix}\K\hat{I}_z.
\end{equation}
Here, $\hat{I}_{xy}$ denotes the inversion of the $k_x$ and $k_y$ coordinates, and $\hat{I}_{z}$ denotes the inversion of the $k_z$ coordinate. $\hat{R}^z_{4\pi/3}$ is the rotation of momenta by $4\pi/3$ through the $k_z$ axis.

The other two IRREPs for $D_3\times \mathbb{Z}_2$ are given by
\begin{equation}
	\pi_+(l)=\begin{bmatrix}
		0 & 1\\
		1 & 0
	\end{bmatrix},\quad \pi_+(m)=\begin{bmatrix}
		1 & 0\\ 
		0 & -1
	\end{bmatrix},\quad \pi_{+}(r)=\begin{bmatrix}
		\cos\frac{2\pi}{3} & i\sin \frac{2\pi}{3}\\
		i\sin \frac{2\pi}{3} & \cos\frac{2\pi}{3} 
	\end{bmatrix},
\end{equation}
and 
\begin{equation}
	\pi_-(l)=\begin{bmatrix}
		0 & 1\\
		1 & 0
	\end{bmatrix},\quad \pi_-(m)=\begin{bmatrix}
		1 & 0\\ 
		0 & -1
	\end{bmatrix},\quad \pi_{-}(r)=\begin{bmatrix}
		\cos\frac{2\pi}{3} & -i\sin \frac{2\pi}{3}\\
		-i\sin \frac{2\pi}{3} & \cos\frac{2\pi}{3} 
	\end{bmatrix}.
\end{equation}
It is easy to see $\pi_{\pm}$ are a complex conjugacy pair. Thus, they give a $4$D IRREPs for the whole little cogroup:
\begin{gather}
	L_z^K=\begin{bmatrix}
		0 & -i & 0 & 0\\
		-i & 0 & 0 & 0\\
		0 & 0 & 0 & -i\\
		0 & 0 & -i & 0
	\end{bmatrix},\quad M=\begin{bmatrix}
		1 & 0 & 0 & 0\\
		0 & -1 & 0 & 0\\
		0 & 0 & 1 & 0\\
		0 & 0 & 0 & -1
	\end{bmatrix}\hat{I}_{y},\quad 
	R=\begin{bmatrix}
		\cos\frac{2\pi}{3} & 0 & 0 & -\sin \frac{2\pi}{3}\\
		0 & \cos\frac{2\pi}{3} & -\sin \frac{2\pi}{3} & 0\\
		0 & \sin \frac{2\pi}{3} & \cos\frac{2\pi}{3} & 0\\
		\sin \frac{2\pi}{3} & 0 & 0 & \cos\frac{2\pi}{3}
	\end{bmatrix}\hat{R}^z_{\frac{4\pi}{3}},\nonumber \\
	R_{\pi}T=\begin{bmatrix}
		1 & 0 & 0 & 0\\
		0 & 1 & 0 & 0\\
		0 & 0 & 1 & 0\\
		0 & 0 & 0 & 1
	\end{bmatrix}\K\hat{I}_z.
\end{gather}
Here, we have used \eqref{Mps2RM}.

To compare with our tight-binding model, it is convenient to use the following matrix representations:
\begin{gather}\label{K_IRREP}
	L_z^K=\begin{bmatrix}
		0 & 0 & 1 & 0\\
		0 & 0 & 0 & 1\\
		-1 & 0 & 0 & 0\\
		0 & -1 & 0 & 0
	\end{bmatrix},\quad M=\begin{bmatrix}
		0 & 1 & 0 & 0\\
		1 & 0 & 0 & 0\\
		0 & 0 & 0 & -1\\
		0 & 0 & -1 & 0
	\end{bmatrix}\hat{I}_{y},\quad 
	R=\begin{bmatrix}
		e^{i2\pi/3} & 0 & 0 & 0\\
		0 & e^{-i2\pi/3} & 0 & 0\\
		0 & 0 & e^{i2\pi/3} & 0\\
		0 & 0 & 0 & e^{-i2\pi/3}
	\end{bmatrix}\hat{R}^z_{\frac{4\pi}{3}},\\
	R_{\pi}T=\begin{bmatrix}
		0 & 1 & 0 & 0\\
		1 & 0 & 0 & 0\\
		0 & 0 & 0 & -1\\
		0 & 0 & -1 & 0
	\end{bmatrix}\K\hat{I}_z.
\end{gather}
which are equivalent to \eqref{L_IRREP} and \eqref{M_IRREP} up to a unitary transformation.

If the mirror symmetry $M_z$ is included, we have the additional algebraic relations:
\begin{equation}
	\{M_z,L_z^K\}=\{M_z,M_y\}=\{M_z,R_\pi T\}=0, \quad M_z^2=1.
\end{equation}
Thus, $M_z$ is represented in the $4$D IRREP as
\begin{equation}\label{Mz_Rep}
	M_z=\begin{bmatrix}
		0 & 0 & 1 & 0\\
		0 & 0 & 0 & 1\\
		1 & 0 & 0 & 0\\
		0 & 1 & 0 & 0
	\end{bmatrix}\hat{I}_z.
\end{equation}

\section{The $k\cdot p$ models}
For a set of matrix representations of symmetry operators, each symmetry $R$ gives an constraint for the $k\cdot p$ model $h(\bm{k})$ as
\begin{equation}
	U_R h(\bm{k}) U_R^\dagger=h(R\bm{k}).
\end{equation}
In practice, we only need to apply the contraints from the generators of the little cogroup. 

The $k\cdot p$ model for the $M$ point of the rectangular model with the IRREP specified by \eqref{L_IRREP_2} is given by
\begin{equation}
	h(\bm{k})=\begin{bmatrix}
		0 & -i\lambda_x k_x & -i\lambda_y k_x & 0\\
		i\lambda_x k_x & 0 & 0 & i\lambda_y k_y\\
		i\lambda_y k_y & 0 & 0 & -i\lambda_x k_x\\
		0 & -i\lambda_y k_y & i\lambda_x & 0
	\end{bmatrix}=\lambda_x k_x \Gamma_2+\lambda_y k_y\Gamma_4 +\mathcal{O}(k^2).
\end{equation}

The $k\cdot p$ model for the $K$ point of the graphite model with the IRREP specified by \eqref{K_IRREP} is given by
\begin{equation}\label{Graphite-kp}
	\begin{split}
		h(\bm{k})=&\begin{bmatrix}
			0 & \lambda_{xy}k_- & -i\lambda_z k_z & -\lambda'_{xy}k_-\\
			\lambda_{xy}k_+ & 0 & \lambda'_{xy}k_+ & i\lambda_z k_z\\
			i\lambda_z k_z & \lambda'_{xy}k_- & 0 & \lambda_{xy}k_-\\
			-\lambda'_{xy}k_+ & -i\lambda_z k_z & \lambda_{xy}k_+ & 0
		\end{bmatrix}+\mathcal{O}(k^2)\\=&\lambda_{xy}(k_x\Gamma^1 + k_y\Gamma^2)+\lambda_z k_z\Gamma_4+\lambda'_{xy}(k_xi\Gamma^1\Gamma^4 + k_yi\Gamma^2\Gamma^4)+\mathcal{O}(k^2).
	\end{split}
\end{equation} 
It is clear that $\lambda'$ terms do not preserve the operator $M_z$ \eqref{Mz_Rep}.

\section{$ PT $ operators of the tight-binding models}

In this section, we derive $PT$ operators of the two tight binding models, which are used to define the topological charges of semimetal phases.

\subsection{Rectangular Model}

Since the time reversal symmetry is preserved under $ \Z_2 $ gauge field, we focus on the spatial inversion symmetry. As shown in Fig.~2(a) of the main text, the original inversion is broken due to the dimerization and $ \Z_2 $ gauge field. Disregarding the $ \Z_2 $ gauge field, we find that there is a glide reflection $ g_x $, the mirror symmetry $ M_y $, and then an off-centered inversion symmetry $ \mathcal{P}=g_xM_y $ with the inversion center as the middle point of $ t $-bond in presence of the dimerization. The glide reflection is given as
\begin{equation}
	g_x=L_yM_x,
\end{equation}
where $ L_y $ is the half translation of the unit vector along $ y $, and $ M_x $ is the mirror reflection as
\begin{equation}
	M_x=\tau_0\otimes\sigma_1\hat{I}_x.
\end{equation}
By Eq.~\eqref{L-k}, $ L_y $ is represented as
\begin{equation}\label{Ly-k}
	L_y(k_y)=\begin{pmatrix}
		0 & 1 \\ e^{ik_y} & 0
	\end{pmatrix}\otimes \sigma_0.
\end{equation}
There is also the mirror symmetry $ M_y $, represented as
\begin{equation}\label{Mxy}
	M_y=\tau_1\otimes\sigma_0\hat{I}_y
\end{equation}
with $ \hat{I}_x( \hat{I}_y )  $ reversing the momentum along $ x(y) $ direction. Then, the off-centered inversion symmetry $ \mathcal{P}=g_xM_y $ is represented as
\begin{equation}
	\mathcal{P}=\begin{pmatrix}
		1 & 0 \\ 0 & e^{ik_y}
	\end{pmatrix}\otimes\sigma_1 \hat{I}, 
\end{equation}
where $\hat{I}=\hat{I}_{x}\hat{I}_{y} $ reverses the momentum. 

Next, we consider the $ \Z_2 $ gauge field with $ \pi $-flux penetrating through each square. The choice of gauge configuration is given in Fig.~2(a) of the main text. In this case, the original off-centered inversion symmetry $ \mathcal{P} $ is violated. However, we can recover it by the gauge transformation $ G $ that reverses the sign of each site in the even rows, which is represented as
\begin{equation}\label{gauge-1}
	G=\tau_3\otimes\sigma_0.
\end{equation}
So, we are led to the $ G $-dressed off-centered inversion symmetry $ P $ as
\begin{equation}\label{P-1}
	P=\begin{pmatrix}
		1 & 0 \\ 0 & -e^{ik_y}
	\end{pmatrix}\otimes\sigma_1\hat{I}.
\end{equation}
Combining Eq.~\eqref{P-1} with the time reversal as $ T=\hat{I}\hat{\mathcal{K}} $, we have the space-time reversal symmetry $ PT $ as
\begin{equation}\label{PT-1}
	PT=\begin{pmatrix}
		1 & 0 \\ 0 & -e^{ik_y}
	\end{pmatrix}\otimes\sigma_1\hat{\mathcal{K}},
\end{equation}
where $ \hat{\mathcal{K}} $ is the complex conjugation. It can be directly checked that $ PT $ satisfies
\begin{equation}
	(PT)^2=1.
\end{equation}

\subsection{Graphite Model}

Following the case of the rectangular model, we first disregard the $ \Z_2 $ gauge field. Due to the dimerization, the twofold rotation $ R_{\pi} $ is violated. However, there is a screwed twofold rotation $ S_{\pi} $ which is the combination of the twofold rotation and the primitive translation along $ z $ direction, $ i.e. $, $ S_{\pi}=L_zR_{\pi} $. By Eq.~\eqref{L-k}, $ L_z $ is represented as
\begin{equation}\label{Lz-k}
	L_z(k_z)=\begin{pmatrix}
		0 & 1 \\ e^{ik_z} & 0
	\end{pmatrix}\otimes \sigma_0.
\end{equation}
The twofold rotation $ R_{\pi} $ is now represented as
\begin{equation}\label{C2z}
	R_{\pi}=\tau_0\otimes\sigma_1\hat{I}_{xy}.
\end{equation}
Combining Eqs.~\eqref{Lz-k} and \eqref{C2z}, we have the screwed twofold rotation as
\begin{equation}\label{S2z}
	S_{\pi}=\begin{pmatrix}
		0 & 1 \\ e^{ik_z} & 0
	\end{pmatrix}\otimes \sigma_1\hat{I}_{xy}.
\end{equation}
There is also the mirror symmetry $ M_z $ to the $ xy $-plane, represented as
\begin{equation}\label{Mz}
	M_z=\tau_1\otimes\sigma_0\hat{I}_z.
\end{equation}
Combining the screwed twofold rotation $ S_{\pi} $ with the mirror symmetry $ M_z $, we have the space inversion $ \mathcal{P}=S_{\pi}M_z $, represented as
\begin{equation}\label{P-2}
	\mathcal{P}=\begin{pmatrix}
		1 & 0 \\ 0 & e^{ik_z}
	\end{pmatrix}\otimes\sigma_1\hat{I},
\end{equation}
where $ \hat{I}$ reverses the momentum. The symmetry $ \mathcal{P} $ is actually the off-centered inversion, which is violated in the presence of $ \Z_2 $ gauge field with the gauge choice shown in Fig.~3(a) of the main text. However, we can recover the original configuration by the gauge transformation $ G $ as Eq.~\eqref{gauge-1}, reversing the sign of each site in even layers. Then, we have the $ G $-dressed spatial inversion symmetry $ P $, represented as
\begin{equation}
	P=\begin{pmatrix}
		1 & 0 \\ 0 & -e^{ik_z}
	\end{pmatrix}\otimes\sigma_1\hat{I}.
\end{equation}
Combining it with the time reversal symmetry $ T=\hat{I}\hat{\mathcal{K}} $, the space-time inversion symmetry $ PT $ is obtained as
\begin{equation}
	PT=\begin{pmatrix}
		1 & 0 \\ 0 & -e^{ik_z}
	\end{pmatrix}\otimes\sigma_1\hat{\mathcal{K}},
\end{equation}
which directly leads to
\begin{equation}
	(PT)^2=1.
\end{equation}

\section{The Dark-Bright Mechanism of Engineering $ \Z_2 $ gauge field}

\begin{figure}[h]
	\centering
	\includegraphics[scale=0.5]{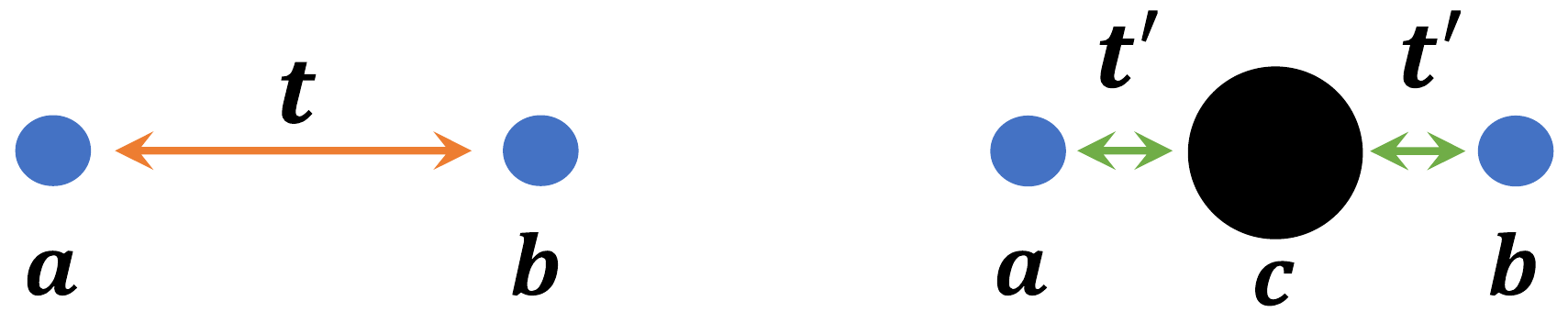}
	\caption{Left one denotes the hopping between two sites. The onsite energy is $ \epsilon $. For the right one, an ancillary site with onsite energy as $ \Delta $ and $ \Delta\gg\epsilon $ is inserted between the two original sites. The hopping energy between original and inserted sites is $ t' $. \label{figs1}}
\end{figure}

Consider two sites with the hopping and onsite energies as $ t>0 $ and $ \epsilon $ in Fig.~\ref{figs1}. 
The Hamiltonian of this system is written as
%\begin{equation}
%H=\epsilon\left(c_a^\dagger c_a+c_b^\dagger c_b\right)+ t\left(c_a^\dagger c_b+c_b^\dagger c_a\right),
%\end{equation}
%by which the Hamiltonian of first quantization can be written as
\begin{equation}
	H=\begin{pmatrix}
		\epsilon & t \\ t & \epsilon
	\end{pmatrix}.
\end{equation}
The eigen state and eigen energy can be obtained as
\begin{equation}
	\begin{split}
		E_+&=\epsilon + t,\qquad   \ket{+}=\ket{a}+\ket{b}, \\
		E_-&=\epsilon - t,\qquad   \ket{-}=\ket{a}-\ket{b},
	\end{split}
\end{equation}
where $ \ket{a}, \ket{b} $ are the local wave functions, or Wannier wave functions. 

For $ t>0 $, the ground state is the anti-bonding state and the excitation is the bonding state. If the sign of $ t $ is reversed as $ -t $, the configuration is exchanged. By inserting an ancillary site between them with onsite energy $ \Delta $ as shown in Fig.~\ref{figs1}, the Hamiltonian is written as
%\begin{equation}
%H=\epsilon\left(c_a^\dagger c_a+c_b^\dagger c_b\right)+\Delta c^\dagger c+ t\left(c_a^\dagger c+c_b^\dagger c +h.c.\right),
%\end{equation}
%by which the Hamiltonian of first quantization is
\begin{equation}
	H'=\begin{pmatrix}
		\epsilon & 0 & t' \\0  & \epsilon& t' \\ t' & t' & \Delta
	\end{pmatrix}.
\end{equation}
%The eigen states and energies can be obtained as
%\begin{equation}
%\begin{split}
%E&=\epsilon \\  % && \phi_a-\phi_b \\
%E&=\frac{\epsilon+\Delta\pm\sqrt{(\Delta-\epsilon)^2+8t'^2}}{2}.
%E&=\frac{\epsilon+\Delta+\sqrt{(\Delta-\epsilon)^2+8t'^2}}{2}
%\end{split}
%\end{equation}
In the limit of $ \Delta \gg \epsilon, t' $, we have the eigen values and vectors as
\begin{equation}
	\begin{split}
		E&=\epsilon, \ \qquad\qquad\qquad  \ket{-}=\left(\ket{a}-\ket{b}\right)/\sqrt{2}, \\
		E&\approx\epsilon-\frac{2t'^2}{\Delta-\epsilon}, \ \qquad  \ket{+}\approx\left( \ket{a}+\ket{b}-\frac{2t'}{\Delta-\epsilon}\ket{c}\right) /\sqrt{2}  ,\\
		E&\approx \Delta+\frac{2t'^2}{\Delta-\epsilon},  \qquad  \ket{e}\approx\frac{t'}{\Delta-\epsilon}\ket{a}+\frac{t'}{\Delta-\epsilon}\ket{b}+\ket{c}.
	\end{split}
\end{equation}
Since $ \Delta\gg \epsilon,t' $, we can take $ \ket{e} $ as high-energy excitation state, which is irrelevant to the energy scale of interest. The state $ \ket{-} $, which is called ``dark state", is decoupled with the inserted site. The state $ \ket{+} $ is called ``bright state". Due to $ \Delta\gg \epsilon,t' $, the occupation on the inserted site can be ignored. Then, in the subspace of dark and bright states as $ \{\ket{-},\ket{+}\} $, we have the Hamiltonian as
\begin{equation}
	H''=\epsilon\ket{-}\bra{-}+\left(\epsilon-\frac{2t'^2}{\Delta-\epsilon}\right)\ket{+}\bra{+} %-\frac{t'^2}{\Delta-\epsilon}\left(\ket{-}\bra{-}+\ket{+}\bra{+}\right). 
\end{equation}
By taking the approximation $ \ket{+}\approx\left(\ket{a}+\ket{b}\right)/\sqrt{2} $ since $ |\frac{2t'}{\Delta-\epsilon}|\ll1 $, we have the effective Hamiltonian in the subspace of $ \{\ket{a},\ket{b}\}$ as
\begin{equation}\label{heff}
	H_{\text{eff}}=\begin{pmatrix}
		\epsilon-\frac{t'^2}{\Delta-\epsilon} & -\frac{t'^2}{\Delta-\epsilon} \\  -\frac{t'^2}{\Delta-\epsilon} &  \epsilon-\frac{t'^2}{\Delta-\epsilon}
	\end{pmatrix},
\end{equation}
which mimics the $ \pi $ hopping phase with the hopping amplitude as $ \frac{t'^2}{\Delta-\epsilon} $. If we set $ t'^2=t(\Delta-\epsilon) $, we have the effective hopping coefficient between the sites $ a $ and $ b $ as $ -t $. Note that the loss of the fidelity comes from the occupation on the inserted site. The higher $ \Delta $ is, the better fidelity the system has.

The two models in the main text can be realized by the dark-bright mechanism as illustrated in Fig.\ref{figs2-0}. It has the advantage that the required alternative dimerization patterns can naturally arise from the approach.

\section{More Band Structures of Tight-binding model}

\begin{figure}[h]
	\centering
	\includegraphics[scale=0.5]{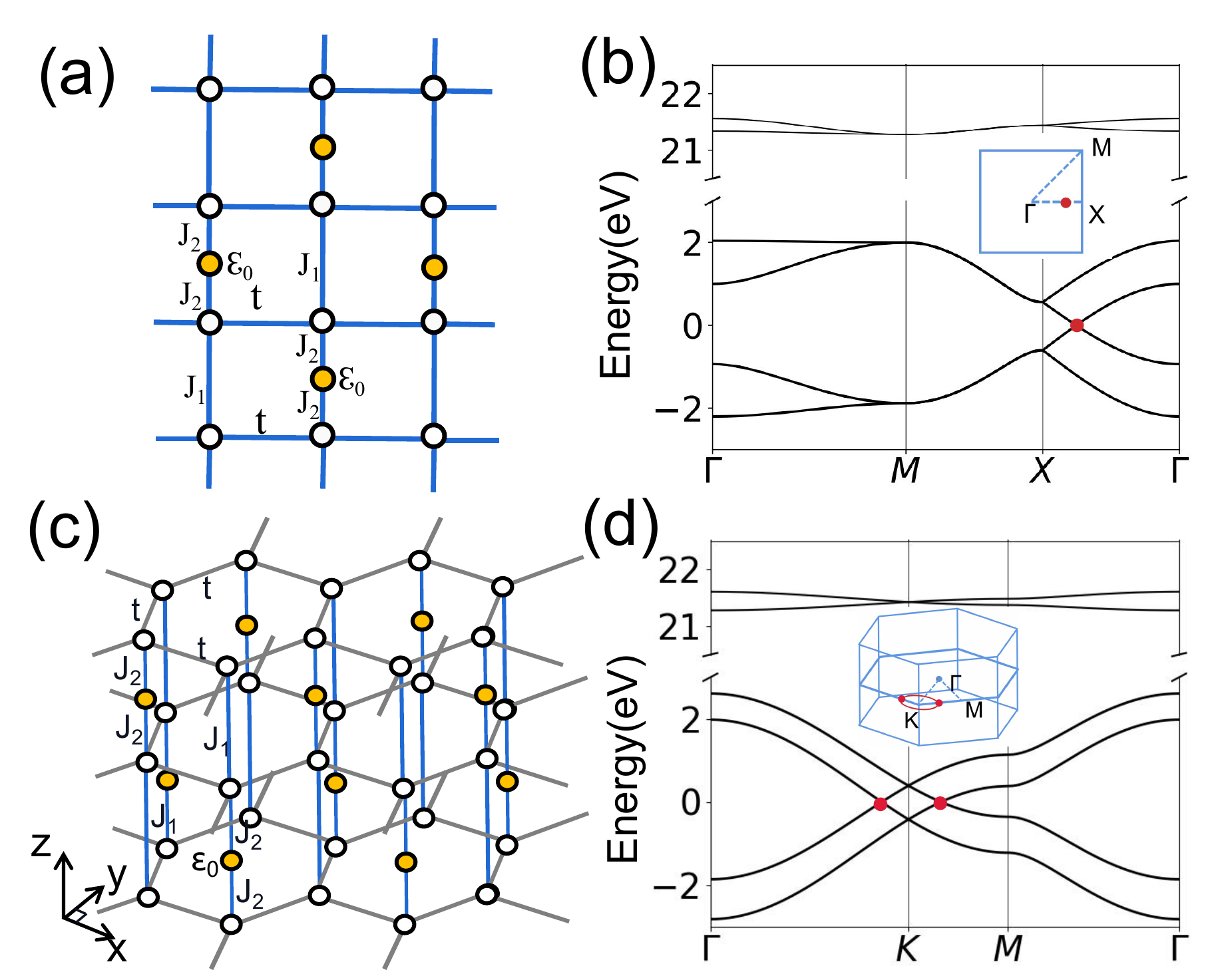}
	\caption{Model realizations by the dark-bright mechanism. (a, c) The high-energy sites (yellow dots) are inserted to a simple rectangular or graphite lattice to obtain effective negative hopping amplitudes. (b) and (d) are the resulting band structures for (a) and (c), respectively. }
	\label{figs2-0}
\end{figure}

In this section, we will give the detailed calculation of the band structures for the two models in the main text, and verify the validity for our strategy of the dark-bright mechanism by comparing the band structures. Fig.~\ref{figs2-0} briefly shows that the low-energy band structures are precisely those for our previously discussed models up to a gauge transformation. Next, we elaborate them one by one.

\begin{figure}[h]
	\centering
	\includegraphics[scale=0.5]{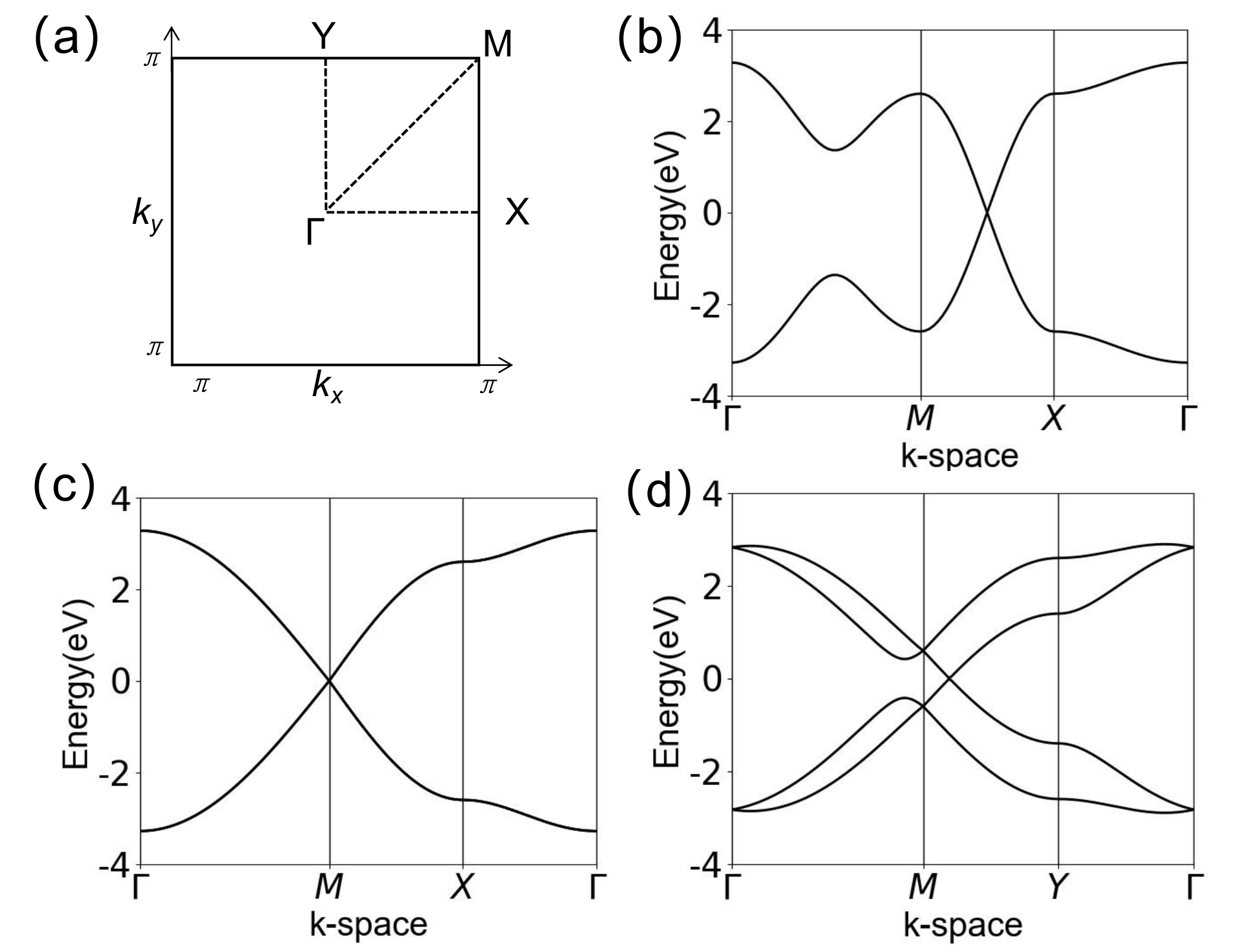}
	\caption{Bulk band structures of our rectangular model. (a) The BZ and the high symmetry lines. (b) The band structure before folding the BZ with $j_1=1.3,j_2=1.3$ in the unit of $ t $. (c) The fourfold degenerate Dirac point at $M=(\pi,\pi)$ after folding the BZ. (d)The splitting of the fourfold degenerate Dirac points into two twofold degenerate Dirac points with $j_1=1.3,j_2=0.7$.}
	\label{figs2}
\end{figure}

The bulk band spectra of our rectangular model are presented in Fig.~\ref{figs2}. Fig.~\ref{figs2}(b) shows the twofold degenerate point at $ (\pi,\pi/2) $, namely, the middle Fermi point of $ MX $, before folding the BZ. There is also a twofold degenerate Fermi point at $ (\pi,-\pi/2) $ due to the time reversal symmetry. After folding the BZ, two Fermi points are mapped to the fourfold degenerate Dirac point at $ M $ as shown in Fig.~\ref{figs2}(c), which is in agreement with our theory of folding BZ in Eq.~(6) of the main text. In presence of the dimerization, the bulk spectrum is shown in Fig.~\ref{figs2}(d) where the degeneracy of Fermi points is decreased to twofold from fourfold, compared with Fig.~\ref{figs2}(c). Note that time reversal symmetry is kept after dimerization. Therefore, there is another Fermi point because of the time reversal symmetry. There is flat boundary band connecting the two Fermi points as shown in Fig.~2(d) of the main text.

\begin{figure}[h]
	\centering
	\includegraphics[scale=0.5]{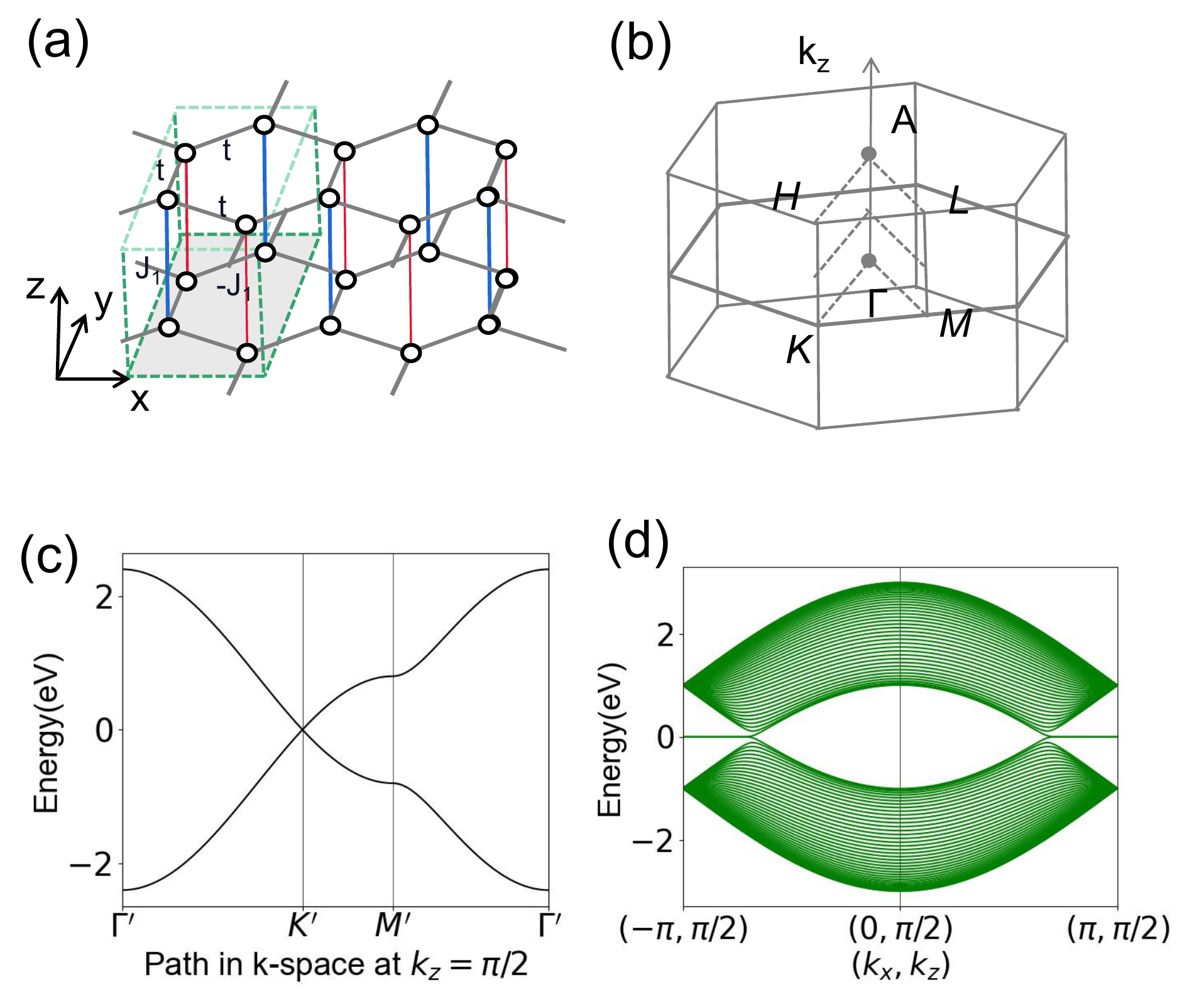}
	\caption{Graphite lattice without folding the BZ along z-direction. (a) The lattice structure where the dashed box denotes the enlarged unit cell in order to obtain the orthogonal basis for the convenience of studying surface states.  (b) BZ of graphite model. (c) Bulk band structure at the plane $k_z=\pi/2$ with twofold degenerate point at $K'=(2\pi/3,\pi/3 )$. We have set $t=0.8,j_1=1.12$. (d) Surface Fermi arcs on the zigzag surface connecting two Weyl points.}
	\label{figs3}
\end{figure}

Without folding the BZ, the band structures of our graphite model are shown in Fig.~\ref{figs3}. The primitive translation along $ z $ is preserved and the BZ is shown in Fig.~\ref{figs3}(b).  In this case, there are twofold degenerate Fermi points at the corners of the plane $ k_z=\pm\pi/2 $ in the BZ, as shown in Fig.~\ref{figs3}(c). Note that the Fermi points at the plane $ k_z=-\pi/2 $ can be derived by time reversal operation on these at the plane $ k_z=\pi/2 $. Without the dimerization, our graphite model is actually a Weyl semimetal, and there are Fermi arcs on the zigzag surface of the graphite crossing the boundary of the BZ and connecting two Weyl points as indicated in Fig.~\ref{figs3}(d). In calculating the surface Fermi arc, we have enlarged the unit cell to obtain an orthogonal basis of the unit vectors as indicated by the dashed box in Fig.~\ref{figs3}(a).

\begin{figure}[t]
	\centering
	\includegraphics[scale=0.45]{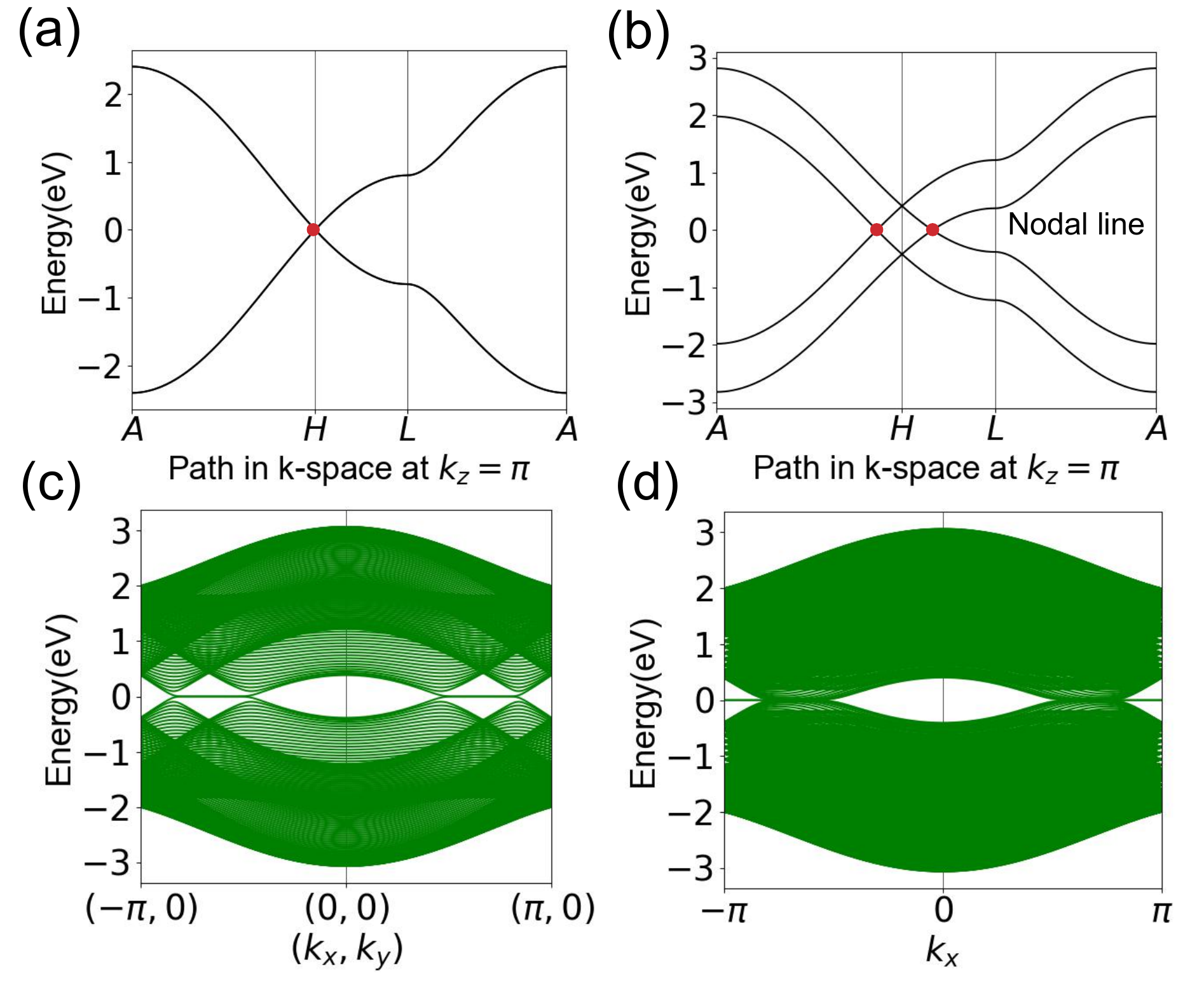}
	\caption{  (a) Bulk band structure with the fourfold degenerate Dirac points by folding the BZ without the dimerization. We have set $t=0.8,j_1=1.12,j_2=1.12$ in the calculation. (b) Bulk band structure with the dimerization. We tune $j_1=0.7$ such that $J_-\neq0$. The red points denote the nodal line intersecting with $ AH $ and $ HL $. (c) The drumhead states on the $ x-y $ surface for a tube geometry. (d) The second-order nodal-line semimetal phase with the hinge Fermi arcs on a pair of opposite hinges along the zigzag direction.}
	\label{figs4}
\end{figure}

\begin{figure}[t]
	\centering
	\includegraphics[scale=0.5]{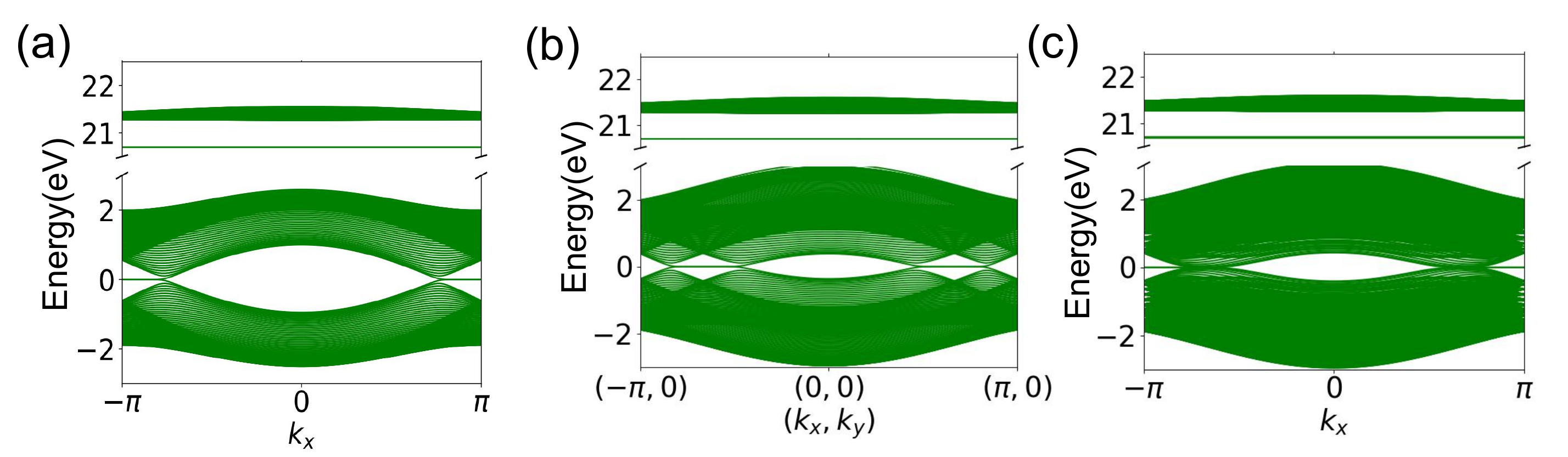}
	\caption{Band structures of tight-binding model by the dark-bright mechanism with the lattice configuration shown in Fig.~4(a) and (c) of the main text. (a) The graphene-like band structure for our rectangle model with $t=1,j_1=1.3,j_2=3.74,\Delta=20$. (b) The spectrum of our graphite with open boundaries perpendicular to $ z $ direction. There are drumhead states inside the projection of the nodal line to $ x-y $ surface. We have set $t=0.8,j_1=1.12,j_2=3.74,\Delta=20$. (c) The spectrum of our graphite model with the open boundaries perpendicular to $ y $ and $ z $ directions.}
	\label{figs5}
\end{figure}

After folding the BZ, the band structures are shown in Fig.~\ref{figs4}. Fig.~\ref{figs4}(a) shows that the twofold degenerate Fermi points are stacked to form fourfold degenerate Dirac points at the corner of the plane $ k_z=\pi $ in the folded BZ by the theory of folding BZ in Eq.~(6) of the main text. By dimerization, the Dirac points are deformed to the nodal lines as discussed in the main text and sketched in Fig.~3(c) of the man text. We show the bulk spectrum along high symmetric lines with dimerization in Fig.~\ref{figs4}(b), where the red points denote the nodal line crossing the lines $ AH $ and $ HL $. Fig.~\ref{figs4}(c) is the spectrum with the open boundary perpendicular to $ z $ direction. There are drumhead states on the surface due to the existence of nodal lines. Fig.~\ref{figs4}(d) presents the spectrum with additional open boundary perpendicular to $ y $ direction and there is hinge Fermi arcs on a pair of opposite hinges as sketched in Fig.~3(d) of the main text.

We verify the validity of the dark-bright mechanism for engineering $ \Z_2 $ gauge field by numerical calculations in Fig.~\ref{figs5}[The details are shown in previous section]. Fig.~\ref{figs5}(a), (b) and (c) are the realistic band spectra of the lattice with the inserted ancillary sites as shown in Fig.~4(a) and (c) of the main text, corresponding to Fig.~2(d) of the main text for our rectangular model, Fig.~\ref{figs4}(c) and (d) for our graphite model, respectively. When the onsite energy of the inserted sites is high enough, the bands of low energy exactly reproduce the ones we have obtained previously except the extra bands of high energy mainly from the inserted lattices. Of course, these extra bands are irrelevant because of their high energy levels. These figures in Fig.~\ref{figs5} demonstrate that the strategy of dark-bright mechanism to engineer the $ \Z_2 $ gauge field is in good agreement with our models.

\section{A brief survey of gauge fields in crystalline systems}
In this section, we give a brief review about simulating $ \Z_2 $ gauge fields in crystalline systems besides the dark-bright mechanism discussed above. 

We first briefly review gauge fields in artificial systems, including cold atoms in optical lattices, photonic/acoustic crystals, periodic mechanical systems, electric circuit arrays below.
\begin{itemize}
	\item In photonic crystals, the gauge field can be generated by modulation of the resonant frequencies, e.g., by adjusting the gap between site ring and link-ring wave guides ~\cite{Ozawa2019,Mittal2019}.
	\item In acoustic crystals, $\mathbb{Z}_2$ hopping phases can be readily realized by coupling the resonators with wave guides on different sides~\cite{Ma2019,Xue2020}.
	\item For cold atoms in optical lattices, we introduce two methods: rotating the optical lattice and laser-assisted tunneling~\cite{Dalibard2011,Zhang2018,Cooper2019}. i) Rotating optical lattice can introduce weak and uniform effective magnetic field and the side effect of Coriolis force should be compensated. ii) For the laser-assisted tunneling, the atomic hopping with desired gauge potentials can be engineered by coupling internal levels of atoms with laser beams. Different kinds of gauge fields can be induced, even the nonabelian ones.
	\item For periodic mechanical systems, effective $\mathbb{Z}_2$ gauge field can be generated by tuning the stiffness coefficients of the spring connections~\cite{Prodan2009}.
	\item For electric circuit arrays, $\mathbb{Z}_2$ gauge fields can be realized by suitably choosing the capacitances and inductances.~\cite{Imhof2018,Yu2020}.
\end{itemize}

For condensed matter systems, we would like to emphasize an important fact, i.e., $ \Z_2 $ gauge fields preserve the time-reversal symmetry, which are essentially different from other $ U(1) $ gauge fields. Thus, $ \Z_2 $ gauge fields can be realized without introducing magnetism or magnetic fields. As such, $ \Z_2 $ gauge fields can be realized in a large class of condensed matter systems with preserved time reversal symmetry. We have discussed the so-called dark-bright mechanism above to achieve $ \Z_2 $ gauge fields. A well-known example is that in cuperates, the effective hopping amplitude between two $Cu$ sites (as mediated by the $O$ site in the middle) is negative.

For strongly correlated systems, there are emergent gauge fields in the low-energy effective theories. The $ \Z_2 $ gauge field, which defines the $ \Z_2 $ spin liquid, can naturally emerge in quantum spin liquid. In the mean-field theory of quantum spin liquid, close to the ground states the spinors are coupled to gauge field, particularly a $ \Z_2 $ gauge field as demonstrated in several works. Actually, perhaps it was the first time that physicists noticed the importance of the projective representations of space groups with a given gauge configuration, which led to Xiao-Gang Wen's theory of PSG classification for quantum phases of spin liquids~\cite{Wen-PSG}. Another example is the Kitaev-type exactly solvable model~\cite{Kitaev2006}, where non-dynamical $ \Z_2 $ gauge fields are coupled with Majorana fermions.

%\bibliographystylesec{plainnat}
%\bibliographysec{Suppl}

%\printbibliography[heading=subbibliography]

%\end{refsection}

%\bibliographystyle{apsrev}
%\bibliography{Suppl}

%\end{appendices}

\end{document}